\documentclass{jpp}
\usepackage{graphicx}
\usepackage{epstopdf, epsfig}
\usepackage{hyperref}

\usepackage{bm}
\newcommand{\be}{\begin{equation}}
\newcommand{\ee}{\end{equation}}
\newcommand{\bea}{\begin{eqnarray}}
\newcommand{\eea}{\end{eqnarray}}

\shorttitle{Dissipation in ABC magnetic fields}
\shortauthor{Q. Chen, K. Nalewajko, B. Mishra}

\title{Scaling of Magnetic Dissipation and Particle Acceleration in~ABC~Fields}

\author{Qiang~Chen\aff{1}
  \corresp{\email{chen@camk.edu.pl}},
  Krzysztof~Nalewajko\aff{1},
  Bhupendra~Mishra\aff{2}}

\affiliation{\aff{1}Nicolaus Copernicus Astronomical Center, Polish Academy of Sciences, Bartycka 18, 00-716 Warsaw, Poland
  \aff{2}Los Alamos National Laboratory, Los Alamos, NM 87545, USA}

\begin{document}

\maketitle

\begin{abstract}
Using particle-in-cell (PIC) numerical simulations with electron-positron pair plasma, we study how the efficiencies of magnetic dissipation and particle acceleration scale with the initial coherence length $\lambda_0$ in relation to the system size $L$ of the two-dimensional (2D) `Arnold-Beltrami-Childress' (ABC) magnetic field configurations.
Topological constraints on the distribution of magnetic helicity in 2D systems,
identified earlier in relativistic force-free (FF) simulations,
that prevent the high-$(L/\lambda_0)$ configurations from reaching the Taylor state,
limit the magnetic dissipation efficiency to about $\epsilon_{\rm diss} \simeq 60\%$.
We find that the peak growth time scale of the electric energy $\tau_{\rm E,peak}$ scales with the characteristic value of initial Alfven velocity $\beta_{\rm A,ini}$ like $\tau_{\rm E,peak} \propto (\lambda_0/L)\beta_{\rm A,ini}^{-3}$.
The particle energy change is decomposed into non-thermal and thermal parts, with non-thermal energy gain dominant only for high initial magnetisation.
The most robust description of the non-thermal high-energy part of the particle distribution is that the power-law index is a linear function of the initial magnetic energy fraction.
\end{abstract}

\begin{keywords}
plasma instabilities, plasma simulation, astrophysical plasmas
\end{keywords}

\section{Introduction}
\label{sec:intro}

Certain high-energy astrophysical sources are characterised by luminous and rapid flares of energetic radiation.
In particular, these include blazars \citep[e.g.,][]{Aharonian_2007,Albert_2007,Aleksi__2011,Abdo_2011,10.1093/mnras/sts711,Ackermann_2016}, and the Crab pulsar wind nebula \citep{Tavani736,Abdo739,Buehler_2012,10.1111/j.1365-2966.2012.22097.x,Mayer_2013,Striani_2013}.
In these extreme astrophysical environments, magnetic fields may dominate even the local rest-mass energy density.
Magnetic reconnection is considered a leading explanation for the efficient particle acceleration behind the dramatic gamma-ray flares of blazars \citep{10.1111/j.1745-3933.2009.00635.x,10.1111/j.1365-2966.2010.18140.x,10.1111/j.1365-2966.2012.21721.x,10.1093/mnras/stt167,10.1093/mnras/stv641,10.1093/mnras/stw1832}.
Through changes of the magnetic line topology, particles are accelerated in the current sheets, converting magnetic energy into kinetic and thermal energy.
In the case of the Crab pulsar wind nebula, the $\gamma$-ray radiation spectral peaks can surpass the classical synchrotron radiation reaction limit ($\sim 160\;{\rm MeV}$),
which suggests a very efficient localised dissipation of magnetic energy that allows for rapid particle acceleration
\citep{Uzdensky_2011,10.1111/j.1365-2966.2011.18516.x,Arons_2012,10.1111/j.1365-2966.2012.21349.x,10.1093/mnras/sts214,B_hler_2014,Zrake_2016,Zrake_2017,lyutikov_komissarov_sironi_porth_2018}.

Numerical simulations based on the kinetic particle-in-cell (PIC) algorithm have demonstrated that relativistic reconnection in collisionless plasma is an efficient mechanism of magnetic energy dissipation and particle acceleration
\citep{Zenitani_2001,doi:10.1063/1.1644814,Zenitani_2007,Lyubarsky_2008,doi:10.1063/1.3589304,Bessho_2012,Kagan_2013,Sironi_2014,PhysRevLett.113.155005,refId0,Guo_2015,Guo_2016,Werner_2016,Werner_2017,10.1093/mnras/stx2530,10.1093/mnras/sty2702,Petropoulou_2019,Guo_2019,guo2020magnetic},
and that it can produce extreme radiative signatures --- energetic, highly anisotropic and rapidly variable \citep{Cerutti_2012,0004-637X-770-2-147,Cerutti_2014,Kagan_2016,nalewajko2018relativistic,10.1093/mnras/sty2636,10.1093/mnras/staa2346,Comisso_2020,10.1093/mnras/staa1899}.
Most of these simulations were initiated from relativistic Harris-type current layers \citep{Kirk_2003}.

An alternative class of magnetostatic equilibria known as the `Arnold-Beltrami-Childress' (ABC) magnetic fields \citep{Arnold_1965} has been recently applied as an
initial configuration for investigating relativistic magnetic dissipation \citep{PhysRevLett.115.095002}.
This configuration involves no kinetically thin current sheets,
but is unstable to the so-called coalescence modes that lead to localised interactions of magnetic domains of opposite polarities, emergence of dynamical current layers, instantaneous particle acceleration, and production of rapid flares of high-energy radiation.
The overall process has been dubbed \emph{magnetoluminescence} -- a generic term for efficient and fast conversion of magnetic energy into radiation \citep{Blandford_2017}.

Numerical simulations of ABC fields have been performed with relativistic magnetohydrodynamics (MHD) and relativistic force-free (FF) algorithms \citep{PhysRevLett.115.095002}.
Detailed comparison between 2D and 3D ABC fields in the FF framework has been performed by \cite{Zrake_East_2016}.
PIC simulations of 2D ABC fields have been reported by \cite{0004-637X-826-2-115} with the focus on the structure of current layers and particle acceleration, by \cite{Yuan_2016} including synchrotron radiation reaction and radiative signatures, and by \cite{nalewajko_yuan_chruslinska_2018} including synchrotron and Inverse Compton (IC) radiation.
ABC fields have been also investigated in great detail (including PIC simulations) by \cite{lyutikov_sironi_komissarov_porth_2017,Lyutikov2017,lyutikov_komissarov_sironi_porth_2018} with application to the Crab Nebula flares.
The first three-dimensional PIC simulations of ABC fields have been reported in \cite{10.1093/mnras/sty2549}.

The previous works have established the following picture.
ABC fields simulated in periodic numerical grids are unstable to coalescence instability if only there exists a state of equal total magnetic helicity and lower total magnetic energy \citep{PhysRevLett.115.095002}.
The growth time scale of the linear coalescence instability is a fraction of the light crossing time scale that depends on the mean magnetisation (or equivalently on the typical Alfven velocity) \citep{0004-637X-826-2-115}.
The magnetic dissipation efficiency is determined primarily by the global magnetic field topology, and it is restricted in 2D systems due to the existence of additional topological invariants \citep{Zrake_East_2016}.
The dissipated magnetic energy is transferred to the particles, resulting in non-thermal high-energy tails of their energy distributions.
These tails can be in most cases described as power laws with a power-law index, but more generally they can be characterised by the non-thermal number and energy fractions \citep{0004-637X-826-2-115}.
With increasing initial magnetisation, the non-thermal tails become harder, containing higher number and energy fractions, similar to the results on Harris-layer reconnection \cite{Sironi_2014,PhysRevLett.113.155005,Werner_2016}.
A limitation of the ABC fields in comparison with the Harris layers is that the initial magnetisation is limited for a given simulation size by the minimum particle densities required to sustain volumetric currents.

The particle acceleration mechanisms of ABC fields, described in more detail in \cite{0004-637X-826-2-115,Yuan_2016,lyutikov_sironi_komissarov_porth_2017}, show similarities to other numerical approaches to the problem of relativistic magnetic dissipation.
During the linear stage of coalescence instability, kinetically thin current layers form and evolve very dynamically.
The few particles that happen to straggle into one of those layers are accelerated by direct non-ideal reconnection electric fields ($\bm{E}\cdot\bm{B} \ne 0$, $|\bm{E}| > |\bm{B}|$).
This is essentially the \cite{Zenitani_2001} picture of magnetic X-point, which is important also in large-scale simulations of Harris-layer reconnection in the sense that particles that pass through a magnetic X-point are most likely to eventually reach top energies \citep{Sironi_2014,Guo_2019}.
The non-linear stage of coalescence instability features slowly damped electric oscillations that gradually convert to particle energies.
This can affect essentially all particles, as electric oscillations cross the entire simulation volume multiple times.
Particles accelerated during the linear stage now propagate on wide orbits and can interact with electric perturbations at random angles.
This is reminiscent of a Fermi process, in particular of the kind envisioned by \cite{2012PhRvL.108m5003H}.
With a larger number of magnetic domains, the coalescence proceeds in multiple stages, with the successive current layers increasingly less regular.
The system becomes chaotic more quickly and begins to resemble a decaying turbulence of the kind studied by \cite{Comisso_2019}.

As the previous PIC simulations of ABC fields were largely limited to the lowest unstable mode, in this work we present the results of new series of 2D PIC simulations of ABC fields for different coherence lengths $\lambda_0$ in order to understand how they affect the efficiency of magnetic dissipation and particle acceleration.
Although the coalescence instability is rather fast, it is followed by slowly damped non-linear oscillations, hence our simulations are run for at least $25 L/c$ light crossing times for the system size $L$ to allow these oscillations to settle.
Our simulations were performed at three different sizes,
in addition we investigated the effects of numerical resolution and local particle anisotropy,
in order to break the relation between the effective wavenumber and the mean initial magnetisation.
We also compare our results with new 3D simulations following the setup described in \cite{10.1093/mnras/sty2549}.

In Section \ref{sec:setup} we define the initial configuration of our simulations.
Our results are presented in Section \ref{sec:res},
including spatial distributions of magnetic fields (Section \ref{sec:mag}),
evolution of the total energy components (Section \ref{sec:evo}),
conservation accuracy of the magnetic helicity (Section \ref{sec:heli}),
and particle energy distributions (Section \ref{sec:acc}).
Discussion is provided in Section \ref{sec:disc}.

\section{Simulation setup}
\label{sec:setup}


We perform a series of PIC simulations using the {\tt Zeltron} code\footnote{\url{http://benoit.cerutti.free.fr/Zeltron/}} \citep{0004-637X-770-2-147} of 2D periodic magnetic equilibria known as ABC fields \citep{PhysRevLett.115.095002}.
As opposed to the Harris layers, these initial configurations do not contain kinetically thin current layers.
In 2D, there are two ways to implement ABC fields on a periodic grid, which we call diagonal or parallel, referring to the orientation of the separatrices between individual magnetic domains.
The \emph{diagonal} ABC field is defined as:
\bea
B_x(x,y) &=& B_0\sin(2\pi y/\lambda_0)\,, \\
B_y(x,y) &=& B_0\cos(2\pi x/\lambda_0)\,, \\
B_z(x,y) &=& B_0\left[\sin(2\pi x/\lambda_0)+\cos(2\pi y/\lambda_0)\right]\,,
\eea
where $\lambda_0$ is the coherence length.
The \emph{parallel} ABC field can be obtained from the diagonal one through rotation by $45^\circ$ and increasing the effective wavenumber by factor $\sqrt{2}$:
\bea
B_x(x,y) &=& B_0\left[\sin(\sqrt{2}\pi(x+y)/\lambda_0)+\sin(\sqrt{2}\pi(x-y)/\lambda_0)\right]/\sqrt{2}\,, \\
B_y(x,y) &=& B_0\left[\sin(\sqrt{2}\pi(x-y)/\lambda_0)-\sin(\sqrt{2}\pi(x+y)/\lambda_0)\right]/\sqrt{2}\,, \\
B_z(x,y) &=& B_0\left[\cos(\sqrt{2}\pi(x+y)/\lambda_0)-\cos(\sqrt{2}\pi(x-y)/\lambda_0)\right]\,.
\eea
With this, both the diagonal and parallel configurations satisfy the Beltrami condition
$\nabla\times\bm{B} = -(2\pi/\lambda_0)\bm{B}$.
In all cases, the mean squared magnetic field strength is $\left<B^2\right> = 2B_0^2$ and the maximum magnetic field strength is $B_{\rm max} = 2B_0$.

These magnetic fields are maintained in an initial equilibrium by volumetric current densities $\bm{j}(\bm{x}) = -(c/2\lambda_0)\bm{B}(\bm{x})$ provided by locally anisotropic particle distribution
\citep[for details, see][]{0004-637X-826-2-115,10.1093/mnras/sty2549}.
ABC fields are characterised by vanishing divergence of the electromagnetic stress tensor $\partial_iT_{\rm EM}^{ij} = 0$ (equivalent to the vanishing $\bm{j}\times\bm{B}$ force),
which implies uniform gas pressure that can be realised with uniform temperature $T$ and uniform gas density $n$.
We chose the initial particle energy distribution to be Maxwell-J\"{u}ttner distribution of relativistic temperature $\Theta = kT/mc^2 = 1$, hence the mean particle energy is $\left<\gamma\right> \simeq 3.37$, and the mean particle velocity is $\left<\beta\right> \simeq 0.906$.
The gas density (including both the electrons and positrons) is given by:
\be
n = \frac{3B_0}{2e \tilde{a}_1 \left<\beta\right> \lambda_0}
\ee
where $\tilde{a}_1 \le 1/2$
is a constant that normalises the dipole moment of the local particle distribution.
We chose $\tilde{a}_1 = 1/4$ as a standard value, but we investigate the effect of reduced local particle anisotropy with lower values of $\tilde{a}_1$ that result in higher particle densities and lower magnetisation values.
The initial kinetic energy density is:
\be
u_{\rm kin,ini} = \left<\gamma\right>nm_{\rm e}c^2 \simeq
\frac{6\pi\left<\gamma\right>}{\tilde{a}_1\Theta\left<\beta\right>}
\left(\frac{\rho_0}{\lambda_0}\right)
\left<u_{\rm B,ini}\right>\,,
\ee
where $\rho_0 = \Theta m_{\rm e} c^2/(eB_0)$ is the nominal gyroradius, and $\left<u_{\rm B,ini}\right> = B_0^2/4\pi$ is the initial mean magnetic energy density.
The initial mean hot magnetisation is given by:
\be
\left<\sigma_{\rm ini}\right> = \frac{\left<B^2\right>}{4\pi w} = \frac{\tilde{a}_1\Theta\left<\beta\right>}{3\pi (\left<\gamma\right>+\Theta)} \left(\frac{\lambda_0}{\rho_0}\right)\,,
\label{eq_sigma_ini}
\ee
where $w = (\left<\gamma\right>+\Theta)nm_{\rm e}c^2$ is the relativistic enthalpy density.
For $\Theta = 1$, we have $\left<\sigma_{\rm ini}\right> \simeq (4\tilde{a}_1)(\lambda_0/182\rho_0)$.

We performed simulations of either diagonal or parallel ABC fields and for different wavenumbers $k$
($k = L/\lambda_0$ for diagonal configuration and $k = L/\sqrt{2}\lambda_0$ for parallel configuration).
For instance, a simulation labelled {\tt diag\_k2} is initiated with a diagonal ABC field with $L/\lambda_0 = 2$.
In order to verify the scaling of our results,
we performed series of simulations for three sizes of numerical grids:
small (s) for $N_x = N_y = 1728$,
medium (m) for $N_x = N_y = 3456$,
and large (l) $N_x = N_y = 6912$.
For numerical resolution $\Delta x = \Delta y = L/N_x$, where $L$ is the physical system size we chose a standard value of $\Delta x = \rho_0/2.4$, but we investigated the effect of increased resolution on the medium numerical grid.
The numerical time step was chosen as $\Delta t = 0.99(\Delta x/\sqrt{2}c)$.
All of our simulations were performed for at least $25 L/c$ light crossing times.
In each case we used 128 macroparticles (including both species) per cell.

We also performed two new 3D simulations for the cases {\tt diag\_k2} and {\tt diag\_k4}, following the configuration described in \cite{10.1093/mnras/sty2549}, but extending them to $25 L/c$. In this case we chose the following parameter values: $N_x = N_y = N_z = 1152$, $\Delta x = \Delta y = \Delta z = \rho_0/1.28$, $\tilde{a}_1 = 0.2$, and 16 macroparticles per cell.

\section{Results}
\label{sec:res}

The key parameters of our large simulations are listed in Table \ref{tab_res}, where we report basic results describing global energy transformations that will be discussed in Section \ref{sec:evo}, and particle energy distributions that will be discussed in Section \ref{sec:acc}.

\begin{table*}
\caption{Global parameters of energy conversion and particle acceleration compared for the 2D and 3D simulations.
The initial values denoted with subscript \emph{ini} are measured at $t=0$, and the final values (\emph{fin}) are averaged over $20 \le ct/L \le 25$.
The initial mean hot magnetisation $\left<\sigma_{\rm ini}\right>$ is computed from Eq. (\ref{eq_sigma_ini}).
The initial magnetic energies $\mathcal{E}_{\rm B,ini}$ are normalised to the total system energy $\mathcal{E}_{\rm tot}$.
The magnetic dissipation efficiency is defined as $\epsilon_{\rm diss} = 1-\mathcal{E}_{\rm B,fin}/\mathcal{E}_{\rm B,ini}$.
We report the peak value $\tau_{\rm E,peak}$ of the linear growth time scale $\tau_E$ of electric energy, which scales like $\mathcal{E}_{\rm E} \propto \exp(ct/L\tau_E)$.
For the final particle energy distributions, we report:
the power law index $p$,
the maximum Lorentz factor $\gamma_{\rm max}$,
and the non-thermal particle energy fraction $f_E$.}
\centering
\begin{tabular}{lrllrccclrc}
\hline\hline
config & $\displaystyle\frac{L}{\lambda_0}$ & $\tilde{a}_1$ & $\displaystyle\frac{\rho_0}{\Delta x}$ & $\left<\sigma_{\rm ini}\right>$ & $\mathcal{E}_{\rm B,ini}$ & $\epsilon_{\rm diss,fin}$ & $\tau_{\rm E,peak}$ & $p$ & $\gamma_{\rm max}$ & $f_E$ \\
\hline
\multicolumn{11}{l}{2D small, $N_x = 1728$} \\
para\_k1 & $  \sqrt{2}$ & 1/4 & 2.4 & 2.8 & 0.65 & 0.26 & 0.25 & 3.1  & 450 & 0.18 \\
para\_k2 & $ 2\sqrt{2}$ & 1/4 & 2.4 & 1.4 & 0.48 & 0.52 & 0.17 & 3.75 & 190 & 0.16 \\
para\_k4 & $ 4\sqrt{2}$ & 1/4 & 2.4 & 0.7 & 0.31 & 0.59 & 0.14 & 4.8  &  60 & 0.07 \\
para\_k8 & $ 8\sqrt{2}$ & 1/4 & 2.4 & 0.4 & 0.19 & 0.65 & 0.17 & ---  &  30 & 0.02 \\
\hline
\multicolumn{11}{l}{2D medium, $N_x = 3456$} \\
para\_k1 & $ \sqrt{2}$ & 1/4 & 2.4 & 5.6 & 0.78 & 0.27 & 0.21 & 2.85 & 870 & 0.31 \\
diag\_k2 &  2          & 1/4 & 2.4 & 4.0 & 0.72 & 0.44 & 0.16 & 2.95 & 620 & 0.34 \\
para\_k2 & $2\sqrt{2}$ & 1/4 & 2.4 & 2.8 & 0.65 & 0.53 & 0.13 & 3.2  & 590 & 0.28 \\ 
diag\_k4 &  4          & 1/4 & 2.4 & 2.0 & 0.56 & 0.57 & 0.11 & 3.65 & 270 & 0.20 \\
para\_k4 & $4\sqrt{2}$ & 1/4 & 2.4 & 1.4 & 0.48 & 0.59 & 0.09 & 3.8  & 190 & 0.15 \\
diag\_k8 &  8          & 1/4 & 2.4 & 1.0 & 0.39 & 0.60 & 0.08 & 4.2  & 100 & 0.10 \\
para\_k8 & $8\sqrt{2}$ & 1/4 & 2.4 & 0.7 & 0.31 & 0.61 & 0.08 & 4.8  & 60  & 0.06
\vspace{1em} \\
para\_k1 & $  \sqrt{2}$ & 1/8  & 2.4 & 2.8 & 0.64 & 0.26 & 0.27 & 3.35 & 320 & 0.18 \\
para\_k1 & $  \sqrt{2}$ & 1/16 & 2.4 & 1.4 & 0.48 & 0.26 & 0.40 & 4.5  &  80 & 0.07 \\
para\_k1 & $  \sqrt{2}$ & 1/32 & 2.4 & 0.7 & 0.31 & 0.25 & 0.68 & ---  &  30 & 0.02 \\
para\_k2 & $ 2\sqrt{2}$ & 1/8  & 2.4 & 1.4 & 0.48 & 0.51 & 0.19 & 4.2  & 150 & 0.13 \\
para\_k2 & $ 2\sqrt{2}$ & 1/16 & 2.4 & 0.7 & 0.31 & 0.48 & 0.34 & ---  &  50 & 0.04 \\
para\_k4 & $ 4\sqrt{2}$ & 1/8  & 2.4 & 0.7 & 0.31 & 0.57 & 0.17 & 5.2  &  60 & 0.05 \\
para\_k4 & $ 4\sqrt{2}$ & 1/16 & 2.4 & 0.4 & 0.19 & 0.54 & 0.46 & ---  &  30 & 0.01 \\
para\_k8 & $ 8\sqrt{2}$ & 1/8  & 2.4 & 0.4 & 0.19 & 0.58 & 0.20 & ---  &  30 & 0.01
\vspace{1em} \\
para\_k1 & $  \sqrt{2}$ & 1/4 &  4.8 & 2.8 & 0.64 & 0.26 & 0.25 & 3.2  & 410 & 0.18 \\
para\_k1 & $  \sqrt{2}$ & 1/4 &  9.6 & 1.4 & 0.48 & 0.26 & 0.32 & 3.8  & 160 & 0.10 \\
para\_k1 & $  \sqrt{2}$ & 1/4 & 19.2 & 0.7 & 0.31 & 0.27 & 0.44 & 5.8  &  40 & 0.05 \\
para\_k2 & $ 2\sqrt{2}$ & 1/4 &  4.8 & 1.4 & 0.48 & 0.52 & 0.17 & 3.75 & 200 & 0.17 \\
para\_k2 & $ 2\sqrt{2}$ & 1/4 &  9.6 & 0.7 & 0.31 & 0.52 & 0.30 & ---  &  50 & 0.10 \\
para\_k4 & $ 4\sqrt{2}$ & 1/4 &  4.8 & 0.7 & 0.31 & 0.58 & 0.13 & 4.75 &  60 & 0.09 \\
para\_k4 & $ 4\sqrt{2}$ & 1/4 &  9.6 & 0.4 & 0.19 & 0.63 & 0.24 & ---  &  30 & 0.05 \\
para\_k8 & $ 8\sqrt{2}$ & 1/4 &  4.8 & 0.4 & 0.19 & 0.64 & 0.15 & ---  &  30 & 0.03 \\
\hline
\multicolumn{11}{l}{2D large, $N_x = 6912$} \\
para\_k1 & $  \sqrt{2}$ & 1/4 & 2.4 & 11.2 & 0.88 & 0.26 & 0.18 & 2.4 & 1490 & 0.56 \\
para\_k2 & $ 2\sqrt{2}$ & 1/4 & 2.4 &  5.6 & 0.78 & 0.53 & 0.10 & 2.95 & 1620 & 0.40 \\
para\_k4 & $ 4\sqrt{2}$ & 1/4 & 2.4 &  2.8 & 0.65 & 0.60 & 0.07 & 3.3 & 510 & 0.26 \\
para\_k8 & $ 8\sqrt{2}$ & 1/4 & 2.4 &  1.4 & 0.48 & 0.61 & 0.05 & 3.85 & 170 & 0.12 \\
\hline
\multicolumn{3}{l}{3D, $N_x = 1152$} \\
diag\_k2 & 2 & 1/5 & 1.28 & 3.6 & 0.71 & 0.50 & 0.22 & 3.2 & 180 & 0.25 \\
diag\_k4 & 4 & 1/5 & 1.28 & 1.8 & 0.54 & 0.75 & 0.17 & 4.0 & 110 & 0.10 \\
\hline
\hline
\end{tabular}
\label{tab_res}
\end{table*}

\subsection{Spatial distribution of magnetic fields}
\label{sec:mag}

\begin{figure*}
\includegraphics[width=\textwidth]{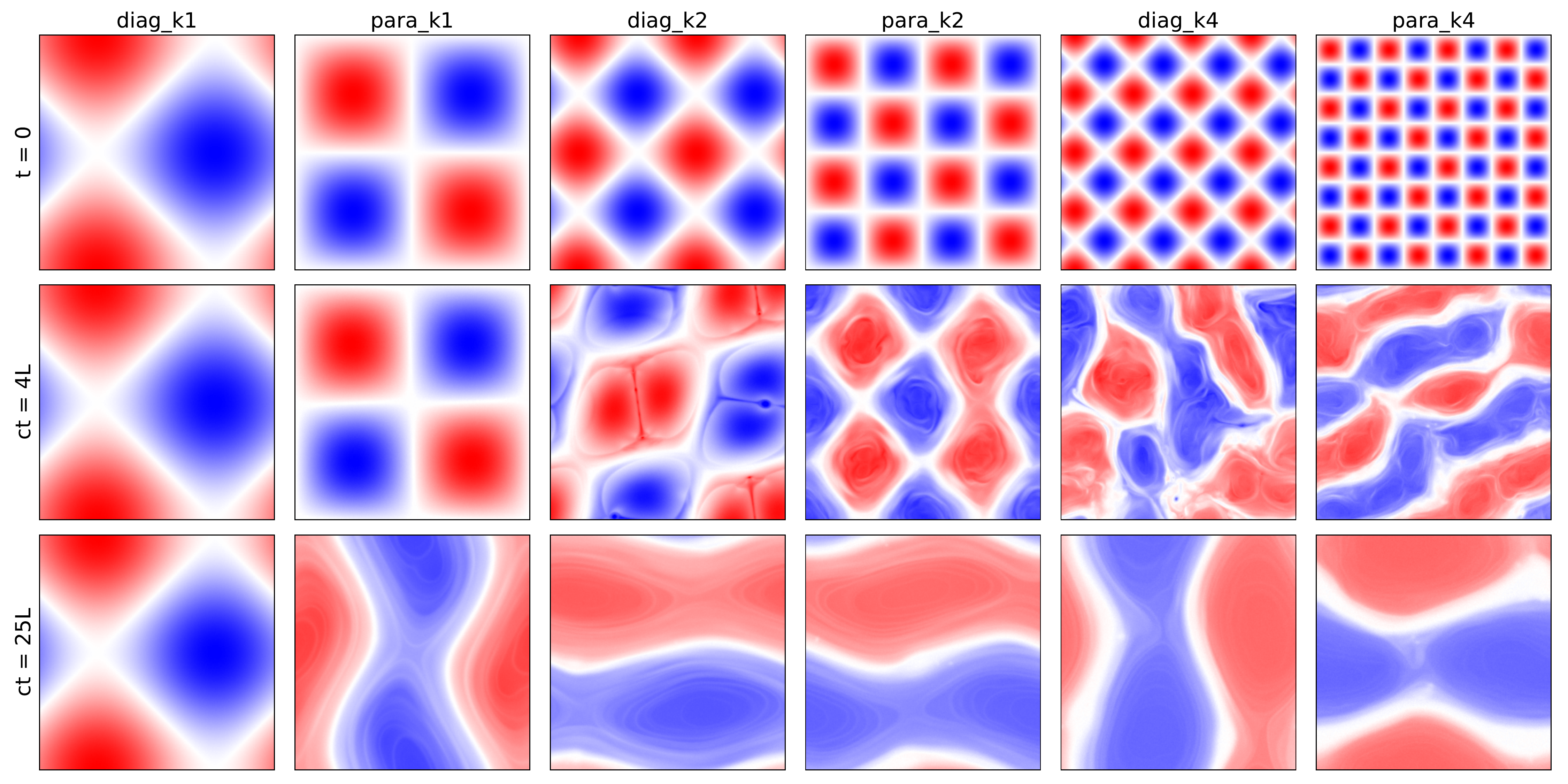}
\caption{Spatial distributions of the out-of-plane magnetic field component $B_z$ for ABC fields of different initial topologies.
Each column of panels compares the initial configuration at $ct/L = 0$ (top) with an intermediate state at $ct/L \simeq 4$ (middle), and with the final state at $ct/L \simeq 25$ (bottom).}
\label{fig_Bz_maps}
\end{figure*}

Fig. \ref{fig_Bz_maps} compares the initial ($ct/L = 0$), intermediate ($ct/L \simeq 4$) and final ($ct/L \simeq 25$) configurations of the out-of-plane magnetic field component $B_z$.
The initial configurations have the form of periodic grids of $B_z$ minima (blue) and maxima (red).
The case {\tt diag\_k1} is the only one that represents a stable equilibrium, as it involves only one minimum and one maximum of $B_z$.
The case {\tt para\_k1} (investigated in detail in \citealt{0004-637X-826-2-115,Yuan_2016}) begins with two minima and two maxima of $B_z$,
by $ct/L \simeq 4$ it is just entering the linear instability stage,
and the final state appears very similar to the case {\tt diag\_k1}, although the domains of positive and negative $B_z$ are still slightly perturbed.
As we increase $L/\lambda_0$, throughout the case of {\tt para\_k4},
the intermediate states become more evolved,
at further stages of magnetic domains coalescence,
while the final states in all cases consist of single positive and negative $B_z$ domains.
We notice that these domains become separated by increasingly broad bands of $B_z \simeq 0$.

\subsection{Total energy transformations}
\label{sec:evo}

The initial configurations investigated here involve various levels of magnetic energy $\mathcal{E}_{\rm B,ini}$ as fractions of the total energy $\mathcal{E}_{\rm tot}$.
The initial magnetic energy fraction decreases with increasing $L/\lambda_0$ and increases with the system size.
Our simulations probe the range of $\mathcal{E}_{\rm B,ini}/\mathcal{E}_{\rm tot}$ values from 0.19 to 0.88.
Related to the initial magnetic energy fraction is the initial mean hot magnetisation $\left<\sigma_{\rm ini}\right>$ (see Eq. \ref{eq_sigma_ini}), which in our simulations takes values from 0.35 to 11.2.

\begin{figure*}
\includegraphics[width=0.497\textwidth]{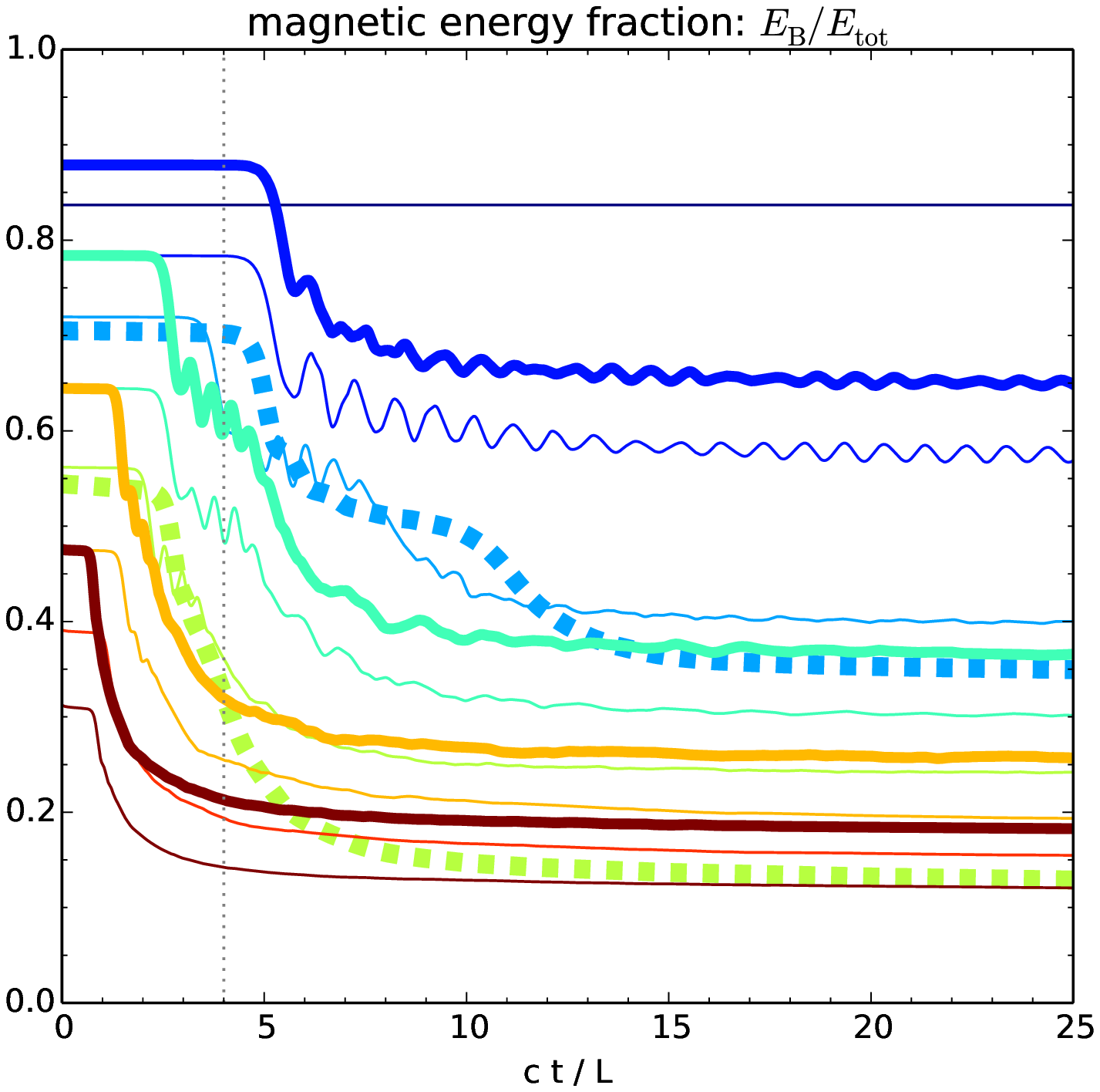}
\includegraphics[width=0.497\textwidth]{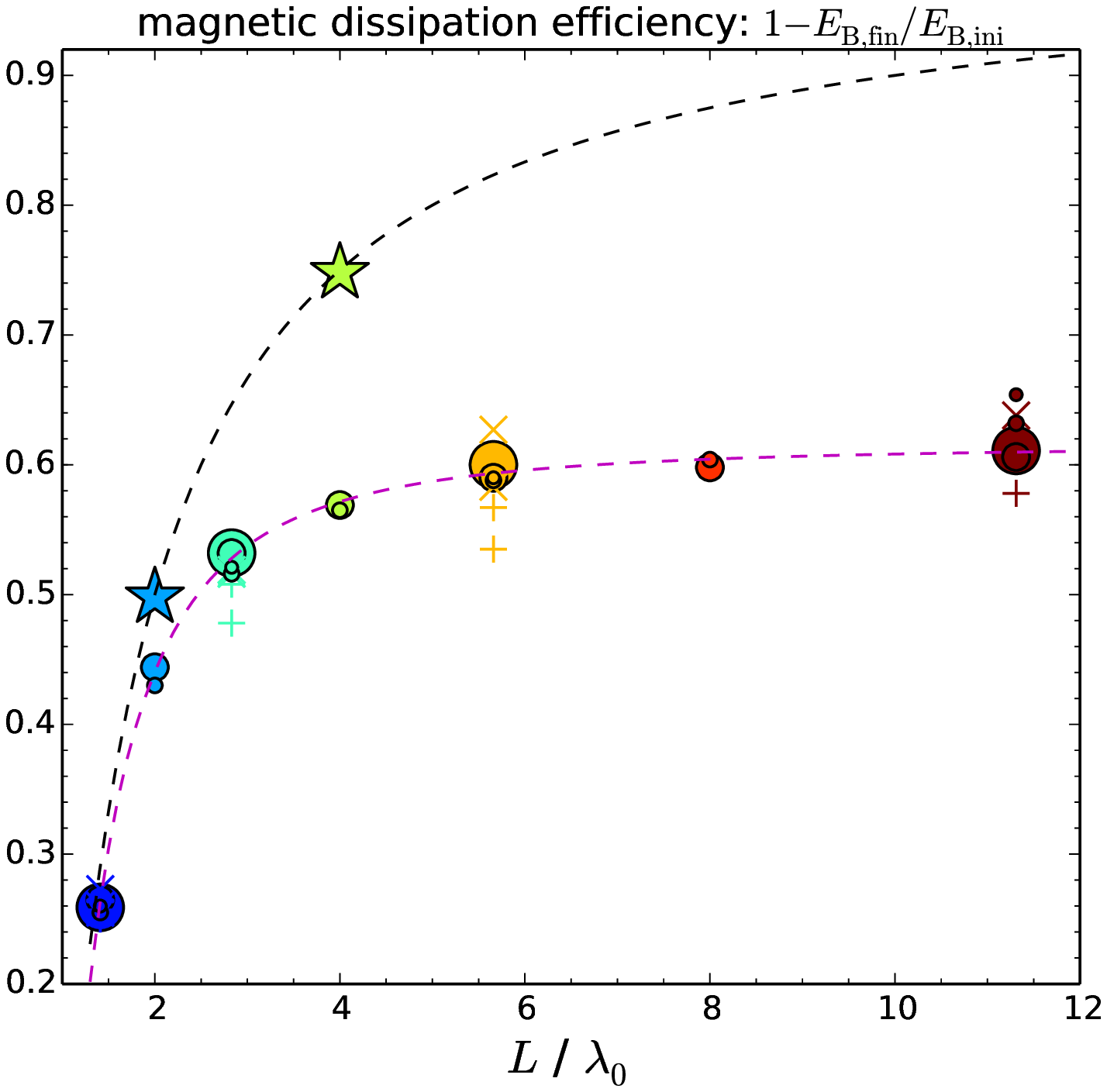}
\caption{\emph{Left panel:} time evolution of the magnetic energy $\mathcal{E}_{\rm B}$ as fraction of the total energy $\mathcal{E}_{\rm tot}$ for the medium (thin solid lines) and large (thick solid lines) simulation sizes.
The thick dashed lines indicate two 3D simulations.
The line colour indicates the effective wavenumber $L/\lambda_0$, as shown in the right panel.
\emph{Right panel:} final magnetic dissipation efficiency $\epsilon_{\rm diss,fin} = 1 - \mathcal{E}_{\rm B,fin} / \mathcal{E}_{\rm B,ini}$ (evaluated at $20 < ct/L < 25$) as function of the effective wavenumber of initial magnetic configuration $L/\lambda_0$.
The large/medium/small circles indicate new results obtained from large/medium/small simulations,
the `+' symbols indicate simulations for non-standard values of $\tilde{a}_1$,
the `x' symbols indicate simulations for non-standard values of $\rho_0/\Delta x$,
and the stars indicate 3D simulations.
The symbol colours indicate the effective wavenumber $L/\lambda_0$.
The black dashed line shows a $1-\lambda_0/L$ relation predicted by the relaxation theorem of \cite{PhysRevLett.33.1139} and matching the 3D results,
and the magenta dashed line shows a $0.62 - 0.70(\lambda_0/L)^2$ relation fitted to the 2D results.}
\label{fig_evo_mag}
\end{figure*}

Time evolutions of the magnetic energy fractions are presented in the left panel of Figure \ref{fig_evo_mag}.
In all studied cases, the magnetic energy experiences a sudden decrease followed by a slow settling.
As the settling is largely complete by $t = 20L/c$, we measure the final magnetic energy fraction $\mathcal{E}_{\rm B,fin}$ as the average over the $20 < ct/L < 25$ period.
We define the final magnetic dissipation efficiency as $\epsilon_{\rm diss,fin} = 1 - \mathcal{E}_{\rm B,fin}/\mathcal{E}_{\rm B,ini}$.
The right panel of Figure \ref{fig_evo_mag} shows that $\epsilon_{\rm diss,fin}$ is a function of magnetic topology parameter $L/\lambda_0$, almost independent of the system size $L$ (although it is slightly lower for reduced values of $\tilde{a}_1$).
For large values of $L/\lambda_0$, magnetic dissipation efficiency appears to saturate at the level of $\epsilon_{\rm diss} \sim 0.6$.
We have fitted the large and medium 2D results for the standard values of $\tilde{a}_1$ and $\rho_0/\Delta x$ with a relation $\epsilon_{\rm diss} = \epsilon_0 - \epsilon_2(\lambda_0/L)^2$, finding $\epsilon_0 \simeq 0.62$ and $\epsilon_2 \simeq 0.70$.

Also shown in Figure \ref{fig_evo_mag} are analogous results for two 3D simulations.
These results are consistent with a relation $\epsilon_{\rm diss} = 1 - \lambda_0/L$ predicted by the relaxation theorem of \cite{PhysRevLett.33.1139}.

\begin{figure*}
\includegraphics[width=0.497\textwidth]{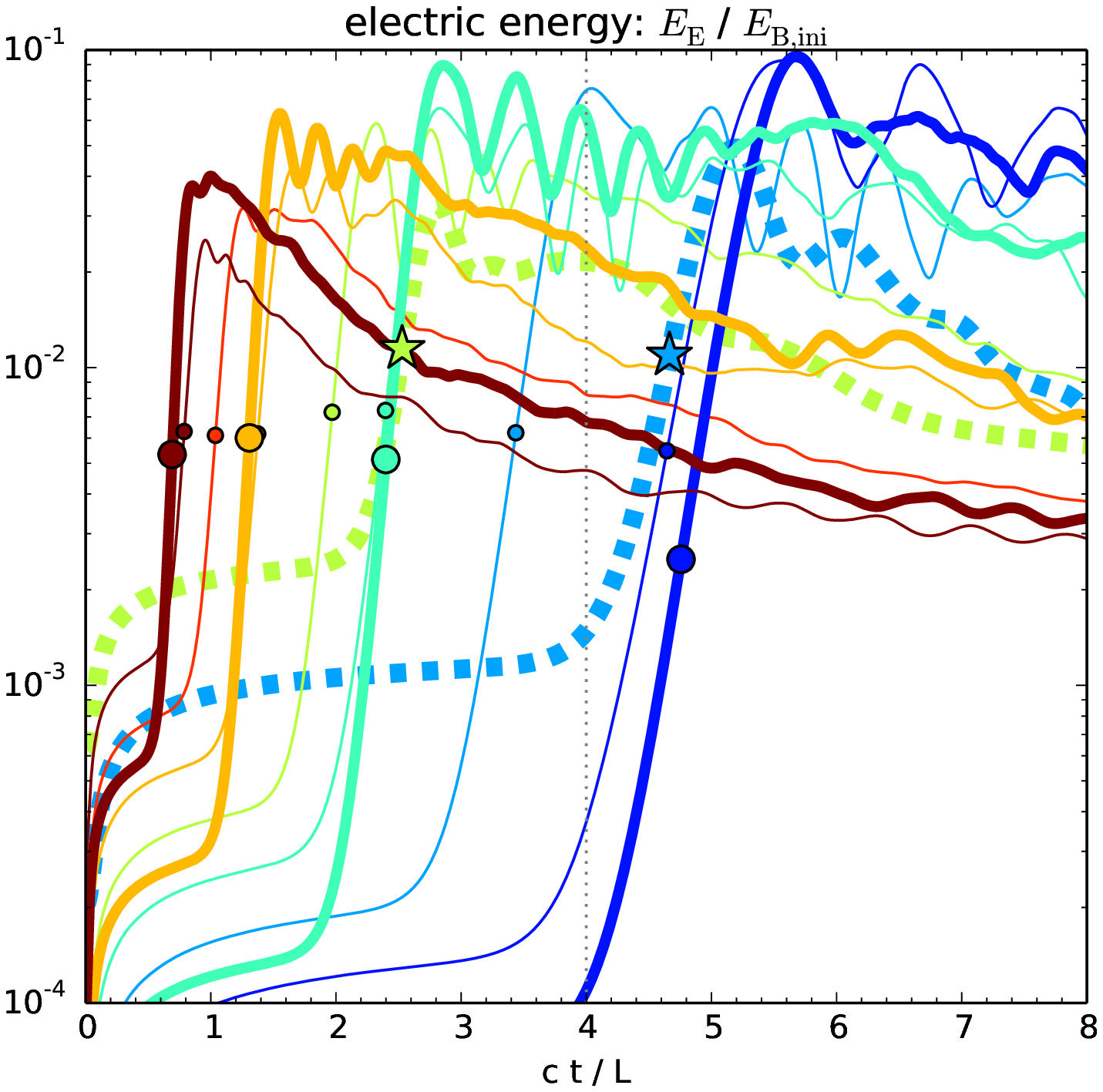}
\includegraphics[width=0.497\textwidth]{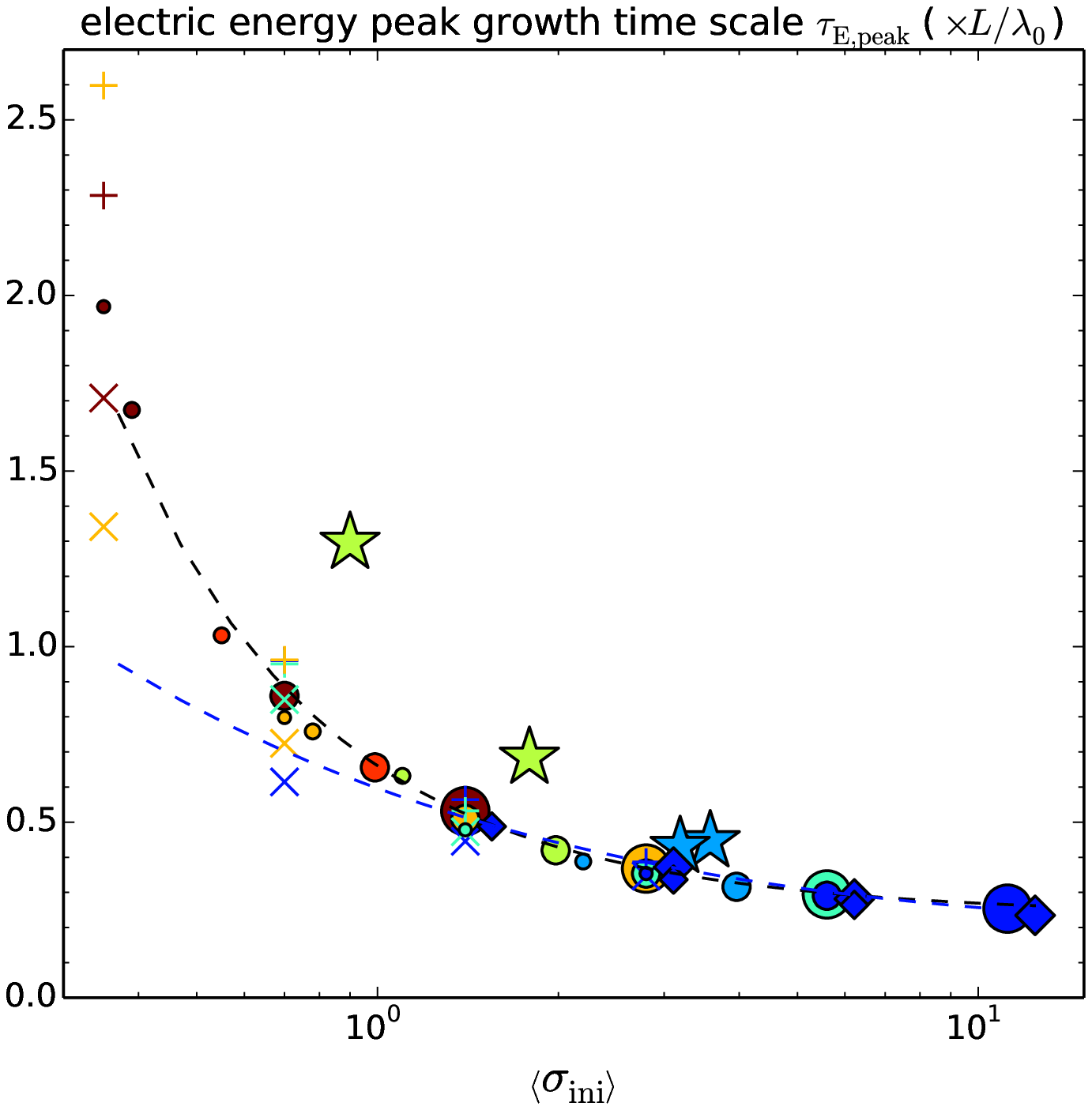}
\caption{\emph{Left panel:} time evolution of the electric energy $\mathcal{E}_{\rm E}$ as fraction of the initial magnetic energy $\mathcal{E}_{\rm B,ini}$.
The line types are the same as in the left panel of Figure \ref{fig_evo_mag}.
Moments of minimum growth time scale are indicated with the filled symbols.
\emph{Right panel:} minimum growth time scales for the total electric energy $\tau_{\rm E}$ as function of the initial mean magnetisation $\left<\sigma_{\rm ini}\right>$.
The symbol types are the same as in the right panel of Figure \ref{fig_evo_mag};
in addition the blue diamonds indicate the {\tt para\_k1} simulations from \cite{0004-637X-826-2-115},
and original shorter 3D runs from \cite{10.1093/mnras/sty2549} are indicated.
The black dashed line shows a $\beta_{\rm A}^{-3}$ trend (see Eq. \ref{eq_tauE_fit}) fitted to all 2D results.
The blue dashed line shows a different trend (see Eq. \ref{eq_tauE_KN16}) suggested previously by \cite{0004-637X-826-2-115}.}
\label{fig_evo_ele}
\end{figure*}

The initial sudden decrease of the magnetic energy is mediated by rapid growth of the electric energy.
Time evolutions of the electric energy $\mathcal{E}_{\rm E}$ as fraction of the initial magnetic energy $\mathcal{E}_{\rm B,ini}$ are presented in the left panel of Figure \ref{fig_evo_ele}.
In all studied cases we find an episode of rapid exponential growth of the electric energy, an indication of linear instability known as coalescence instability \citep{PhysRevLett.115.095002}.
We indicate moments of peak electric energy growth time scale $\tau_{\rm E,peak}$ (defined by $\mathcal{E}_{\rm E} \propto \exp(ct/L\tau_{\rm E})$).
The right panel of Figure \ref{fig_evo_ele} compares the values of $\tau_{\rm E,peak}$, multiplied by $L/\lambda_0$, as function of the initial mean magnetisation $\left<\sigma_{\rm ini}\right>$.
Combining our 2D results with the previous simulations for the case {\tt para\_k1} reported in \cite{0004-637X-826-2-115},
the relation between $\tau_{\rm E,peak}$ and $\left<\sigma_{\rm ini}\right>$ for the standard values of $\tilde{a}_1$ and $\rho_0/\Delta x$ has been fitted as:
\be
\tau_{\rm E,peak} \simeq \frac{0.233\pm 0.005}{(L/\lambda_0)\beta_{\rm A,ini}^3}\,,
\label{eq_tauE_fit}
\ee
where $\beta_{\rm A,ini} = [\left<\sigma_{\rm ini}\right>/(1+\left<\sigma_{\rm ini}\right>)]^{1/2}$ is the characteristic value of initial Alfven velocity.
The four 3D simulations (including two new full runs and two shorter runs from \citealt{10.1093/mnras/sty2549}) show longer growth time scales compared with their 2D counterparts, with the cases {\tt para\_k4} being strongly affected by the noise component of the electric field.

\subsection{Conservation of total energy and magnetic helicity}
\label{sec:heli}

\begin{figure}
\includegraphics[width=0.495\columnwidth]{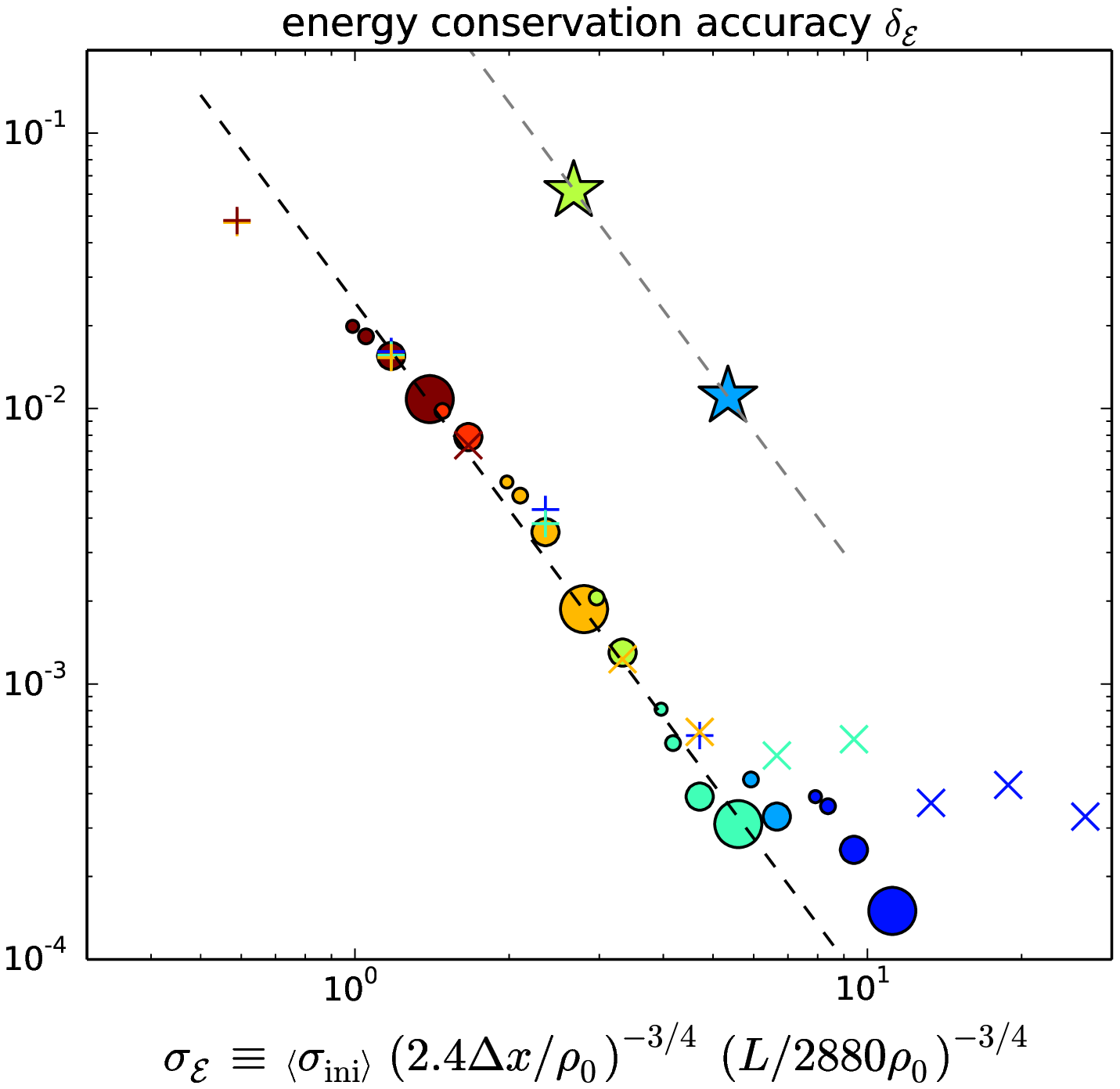}
\includegraphics[width=0.495\columnwidth]{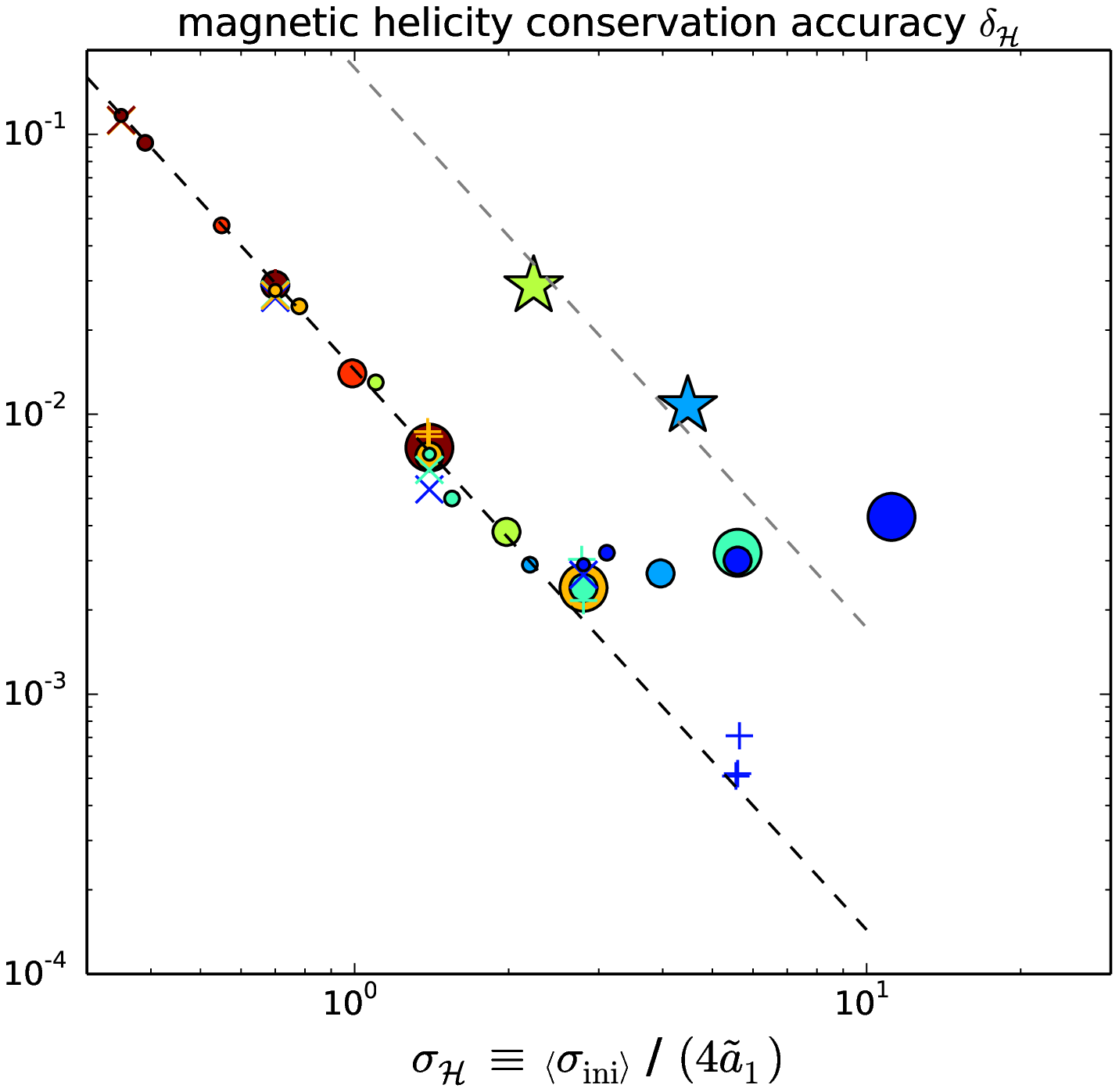}
\caption{Conservation accuracies of total energy $\delta_\mathcal{E}$ (\emph{left panel}) and total magnetic helicity $\delta_\mathcal{H}$ (\emph{right panel}) as functions of modified magnetisation parameters $\sigma_\mathcal{E}$ and $\sigma_\mathcal{H}$, respectively, chosen to minimise scatter around the suggested trends (dashed lines; $\sigma_\mathcal{E}^{-5/2}$ and $\sigma_\mathcal{H}^{-2}$, respectively), see the main text for details.
The symbol types are the same as in the right panel of Figure \ref{fig_evo_mag}.}
\label{fig_cons}
\end{figure}

Figure \ref{fig_cons} shows the conservation accuracy for the total system energy $\mathcal{E}_{\rm tot}$ and total magnetic helicity $\mathcal{H} = \int H\,{\rm d}V$ (where $H = \bm{A}\cdot\bm{B}$ with $\bm{A}$ the magnetic vector potential).
The conservation accuracy for parameter $X$ is defined as $\delta_X \equiv \max|X(ct<25L)/X(t=0)-1|$.
The conservation accuracy of total energy $\delta_\mathcal{E}$ is presented as function of modified magnetisation parameter $\sigma_\mathcal{E} \equiv \left<\sigma_{\rm ini}\right> (2.4\Delta x/\rho_0)^{-3/4} (L/2880\rho_0)^{-3/4}$.
For $1 < \sigma_\mathcal{E} < 6$ (essentially for $L/\lambda_0 \gtrsim 2\sqrt{2}$), energy conservation accuracy scales like $\delta_\mathcal{E} \propto \sigma_\mathcal{E}^{-5/2} \propto \left<\sigma_{\rm ini}\right>^{-5/2} (\Delta x/\rho_0)^{15/8 \simeq 2} (L/\rho_0)^{15/8 \simeq 2}$,
reaching the value of $\simeq 0.02$ for $\sigma_\mathcal{E} \simeq 1$.
For $\sigma_\mathcal{E} > 6$, energy conservation accuracy is found to be of the order $\delta_\mathcal{E} \sim 3\times 10^{-4}$.
In the 3D cases, energy conservation is found to be worse by factor $\simeq 30$ as compared with the 2D results for the same value of $\sigma_\mathcal{E}$.

The conservation accuracy of total magnetic helicity $\delta_\mathcal{H}$ is presented as function of a different modified magnetisation parameter $\sigma_\mathcal{H} \equiv \left<\sigma_{\rm ini}\right> / (4\tilde{a}_1) \simeq \lambda_0/182\rho_0$ (the latter assuming $\Theta=1$).
For $\sigma_\mathcal{H} < 2.5$, magnetic helicity conservation accuracy scales like $\delta_\mathcal{H} \propto \sigma_\mathcal{H}^{-2} \propto (\lambda_0/\rho_0)^{-2}$, reaching the value of $\simeq 0.1$ for $\sigma_\mathcal{H} \simeq 0.4$.
For $\sigma_\mathcal{H} > 2.5$ (essentially for $L/\lambda_0 \lesssim 2$), we find that simulations with reduced values of $\tilde{a}_1$ appear to follow the same trend, however large and medium simulations with standard $\tilde{a}_1$ value show worse conservation of the order $\delta_\mathcal{H} \sim 3\times 10^{-3}$.
In the 3D cases, magnetic helicity conservation is found to be worse by factor $\sim 12$ as compared with the 2D results for the same value of $\sigma_\mathcal{H}$.

\subsection{Particle energy distributions}
\label{sec:acc}

\begin{figure}
\center
\includegraphics[width=0.7\columnwidth]{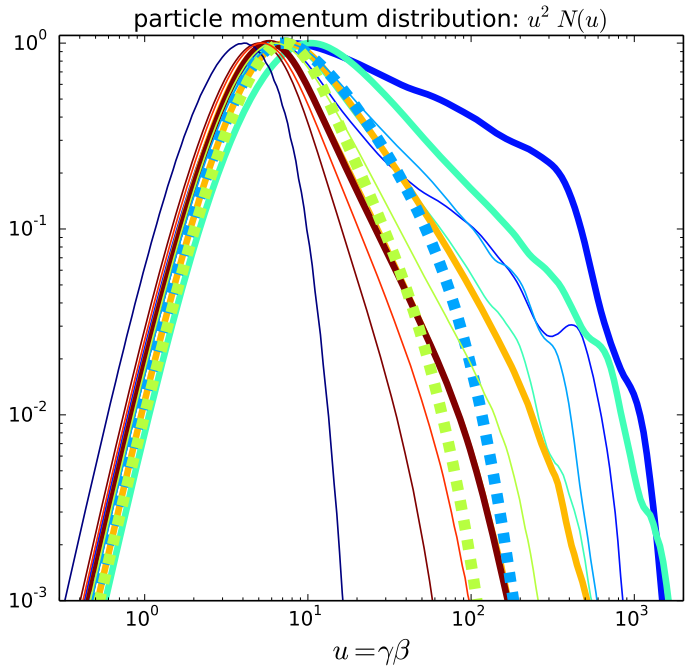}
\caption{Momentum distributions $u^2 N(u)$ of electrons and positrons averaged over the time period $20 < ct/L < 25$.
The line types are the same as in the left panel of Figure \ref{fig_evo_mag}.
}
\label{fig_spec}
\end{figure}

Figure \ref{fig_spec} shows the particle momentum distributions $N(u)$ (closely related to the energy distributions for $u = \sqrt{\gamma^2-1} \gg 1$) for the final states of the medium and large 2D simulations, as well as the 3D simulations (averaged over the time range of $20 < ct/L < 25$).
The non-evolving case {\tt diag\_k1} is equivalent to the initial Maxwell-J\"{u}ttner distribution.
A high-energy excess is evident in all other cases.

There are several ways to characterise this excess component.
In most cases, a power-law section can be clearly identified.
Accurate evaluation of the corresponding power-law index $p$ (such that $N(u) \propto u^{-p}$) is in general complicated, as it requires fitting analytical functions that properly represent the high-energy cutoff \citep{Werner_2016}.
Here, in order to avoid those complications, we estimate a power-law index using a compensation method, multiplying the measured distribution by $u^p$ with different $p$ values to obtain the broadest and most balanced plateau section.
The accuracy of this method is estimated at $\pm 0.05$.
The best values of $p$ estimated for our simulations are reported in Table \ref{tab_res}.
No power-law sections could be identified for certain cases with low initial magnetisations $\left<\sigma_{\rm ini}\right> < 1$.
The hardest spectrum with $p \simeq 2.4$ has been found for the large simulation {\tt para\_k1}.
A similar spectrum with $p \simeq 2.45$ (reexamined with the same method) has been obtained in previous simulations for the case {\tt para\_k1} reported in \cite{0004-637X-826-2-115} and characterised by slightly higher initial magnetisation of $\left<\sigma_{\rm ini}\right> = 12.4$.

\begin{figure*}
\includegraphics[width=0.497\textwidth]{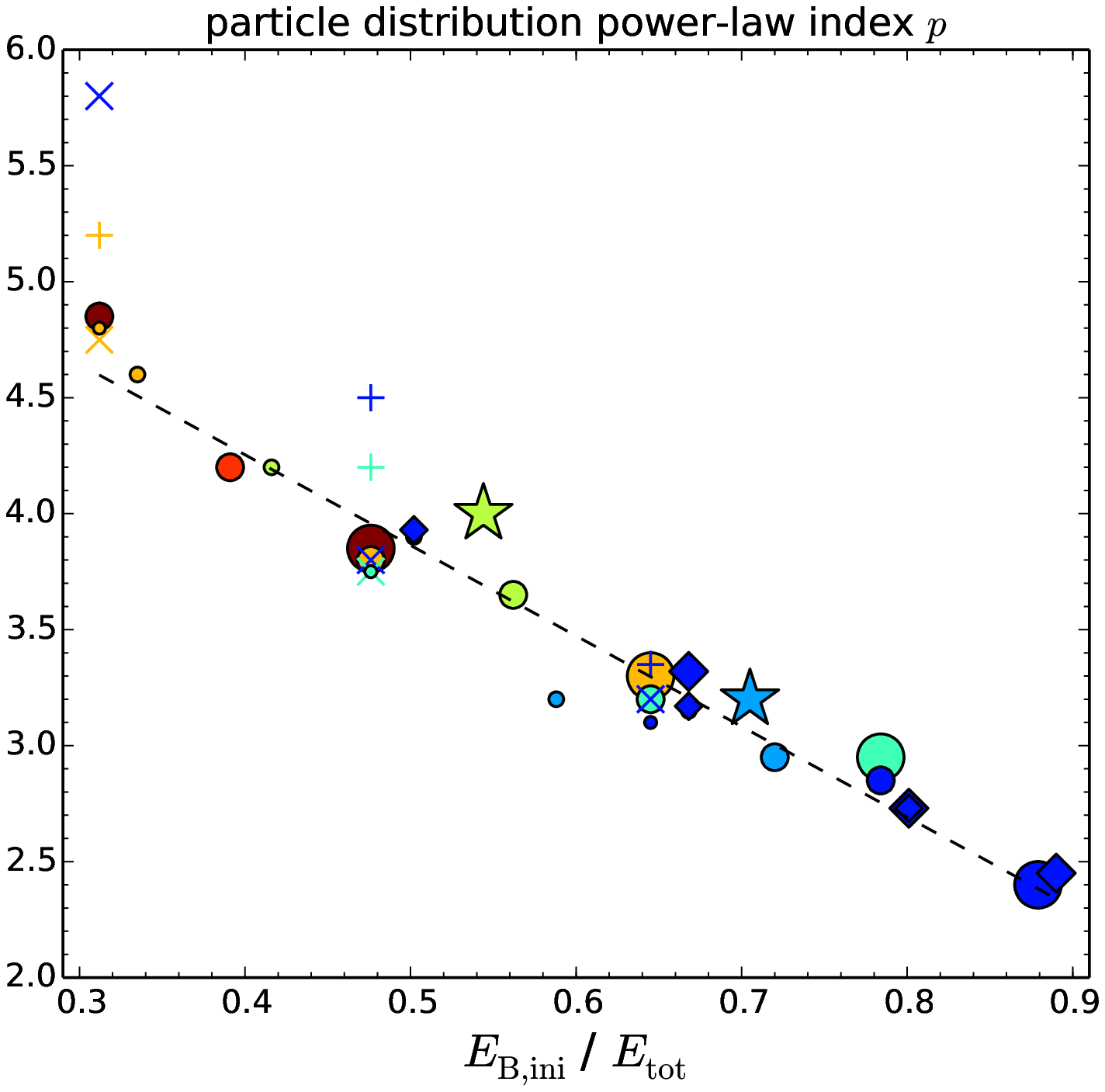}
\includegraphics[width=0.497\textwidth]{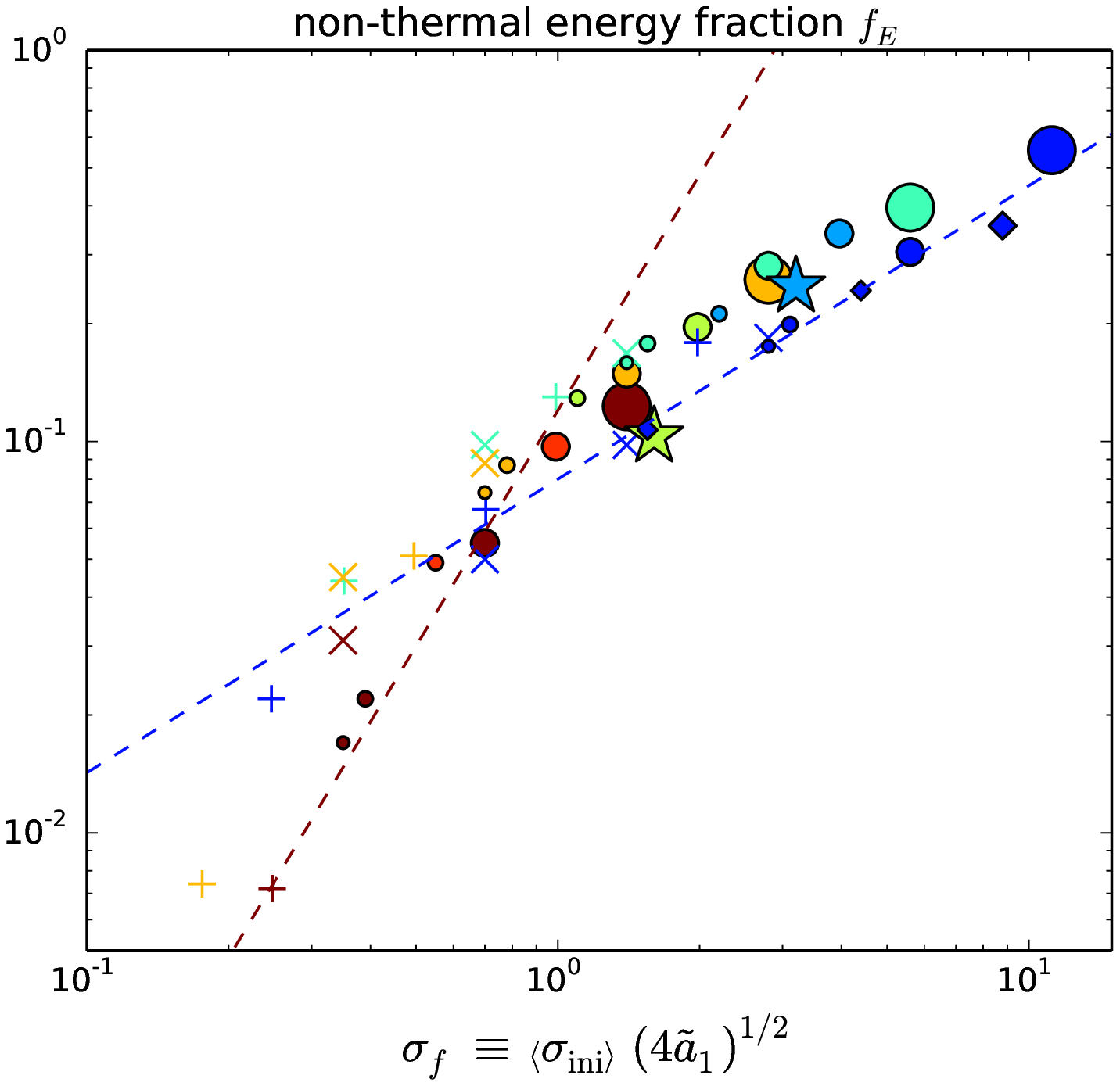}
\caption{\emph{Left panel:} power-law index $p$ of the momentum distribution $N(u) \propto u^{-p}$ as function of the initial magnetic energy fraction $\mathcal{E}_{\rm B,ini}/\mathcal{E}_{\rm tot}$.
The black dashed line shows a linear trend fitted to all 2D results.
\emph{Right panel:} non-thermal energy fraction $f_E$ as function of a modified magnetisation parameter $\sigma_f$.
The dashed lines indicate two trends: $\propto \sigma_f^{3/4}$ (blue) and $\propto \sigma_f^{2}$ (brown).
For both panels, the symbol types are the same as in the right panel of Figure \ref{fig_evo_ele}.}
\label{fig_spec_stat}
\end{figure*}

The left panel of Figure \ref{fig_spec_stat} shows the power-law index $p$ as function of the initial magnetic energy fraction $\mathcal{E}_{\rm B,ini}/\mathcal{E}_{\rm tot}$.
The value of $p$ is strongly anti-correlated with $\mathcal{E}_{\rm B,ini}/\mathcal{E}_{\rm tot}$,
independent of the simulation size,
with the Pearson correlation coefficient of $\simeq -0.98$.
A linear trend has been fitted to the results of 2D simulations with standard values of $\tilde{a}_1$ and $\rho_0/\Delta x$, including the previous {\tt para\_k1} simulations from \cite{0004-637X-826-2-115}:
\be
p \simeq (-3.9 \pm 0.2) \frac{\mathcal{E}_{\rm B,ini}}{\mathcal{E}_{\rm tot}} + (5.8 \pm 0.1) \,.
\label{eq_p_fit}
\ee
Also shown are results for two 3D simulations showing particle distributions slightly steeper as compared with 2D simulations with comparable initial magnetic energy fractions.

The high-momentum excess component of the particle distribution can be alternatively characterised by the maximum particle energy reached $\gamma_{\rm max}$.
Here, the value of $\gamma_{\rm max}$ is evaluated at the fixed level of $10^{-3}$ of the $u^2 N(u)$ distribution normalised to peak at unity (cf. the bottom edge of Figure \ref{fig_spec}).
The final values of $\gamma_{\rm max}$ for our large simulations are reported in Table \ref{tab_res}.
The highest value of $\gamma_{\rm max} \simeq 1620$ has been found for the large simulation {\tt para\_k2}.
For the cases where the power-law index $p$ could be evaluated (note that $\gamma_{\rm max}$ can always be evaluated), $\log\gamma_{\rm max}$ is strongly anti-correlated with $p$, with the Pearson correlation coefficient of $\simeq -0.99$.

Yet another approach to the high-momentum excess is to fit and subtract a low-momentum Maxwell-J\"{u}ttner component and to calculate the non-thermal fractions of particle number $f_n$ and particle energy $f_E$ contained in the remaining excess.
This fitting was performed using the weighted least squares method with the weights proportional to $u^{-2}$.
In all cases, the non-thermal number fractions were found to be closely related to the energy fractions as $f_n \simeq f_E/3.5$.
The values of non-thermal energy fractions $f_E$ for our simulations are reported in Table \ref{tab_res}.
The highest value of $f_E \simeq 56\%$ has been found for the large simulation {\tt diag\_k1}.
For the cases where $p$ could be evaluated, $f_E$ is anti-correlated with $p$, with the Pearson correlation coefficient of $\simeq -0.93$.

The right panel of Figure \ref{fig_spec_stat} shows the non-thermal energy fraction $f_E$ vs. another modified magnetisation parameter $\sigma_f \equiv \left<\sigma_{\rm ini}\right> (4\tilde{a}_1)^{1/2}$.
We also indicate the $f_E \propto \left<\sigma_{\rm ini}\right>^{3/4}$ trend suggested by \cite{0004-637X-826-2-115} and re-fitted only to the {\tt para\_k1} results (deep blue symbols).
We confirm that this trend describes the {\tt para\_k1} results reasonably well, however, it is not followed by the high-$(L/\lambda_0)$ cases that probe lower magnetisation values $\sigma_f < 1$.
In the particular case of $L/\lambda_0 = 8\sqrt{2}$ (brown symbols), the values of $f_E$ decrease faster with decreasing $\sigma_f$, roughly like $f_E \propto \sigma_f^{2}$ for $\sigma_f < 1$.
For intermediate magnetisation values $1 < \sigma_f < 10$, the values of $f_E$ for $L/\lambda_0 > \sqrt{2}$ are systematically higher as compared with the {\tt para\_k1} trend line.
The 3D simulations produced $f_E$ values that are consistent (in the case {\tt diag\_k2}) or somewhat lower (in the case {\tt diag\_k4}) than the 2D results.

\begin{figure}
\includegraphics[width=0.495\columnwidth]{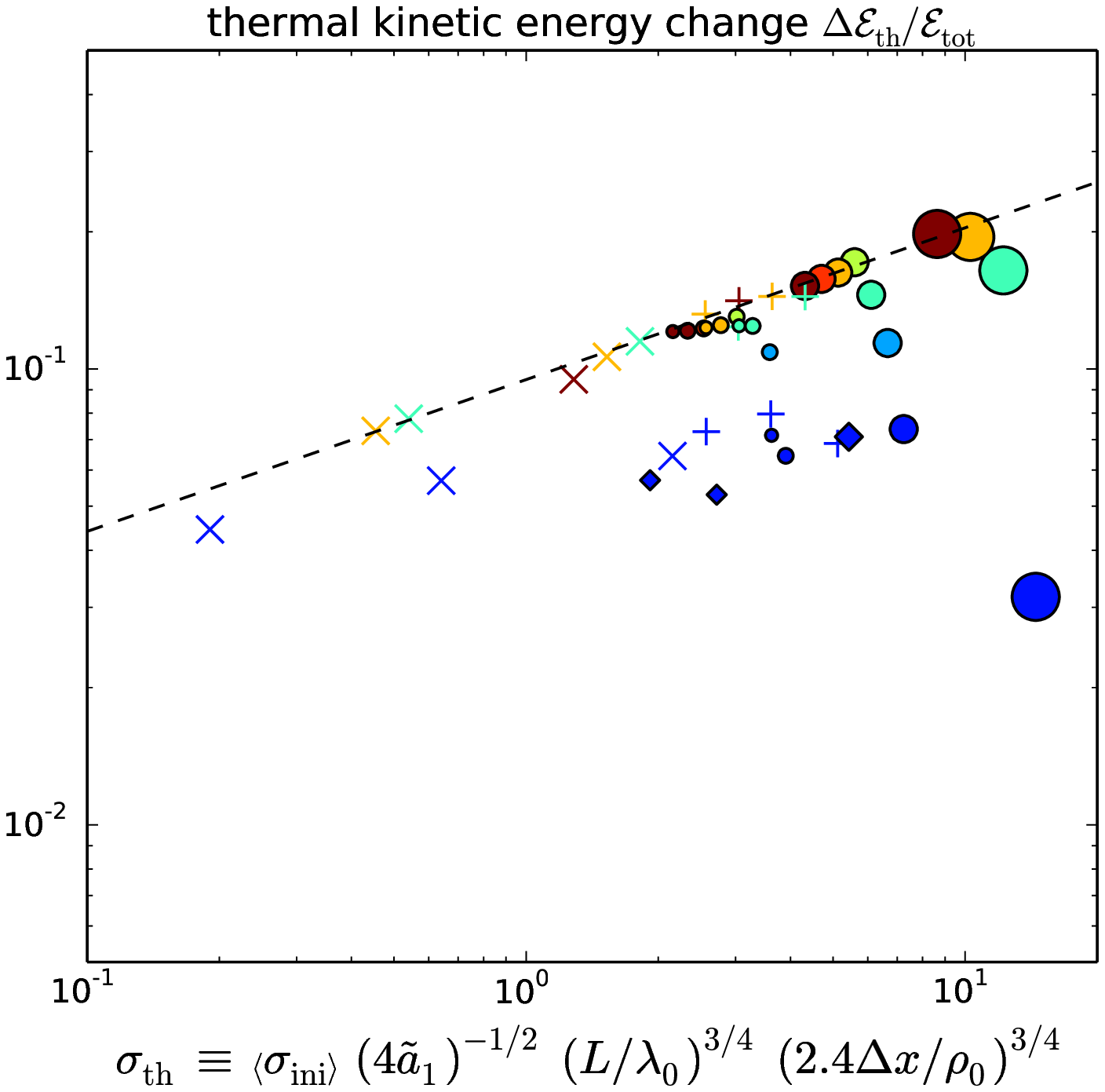}
\includegraphics[width=0.495\columnwidth]{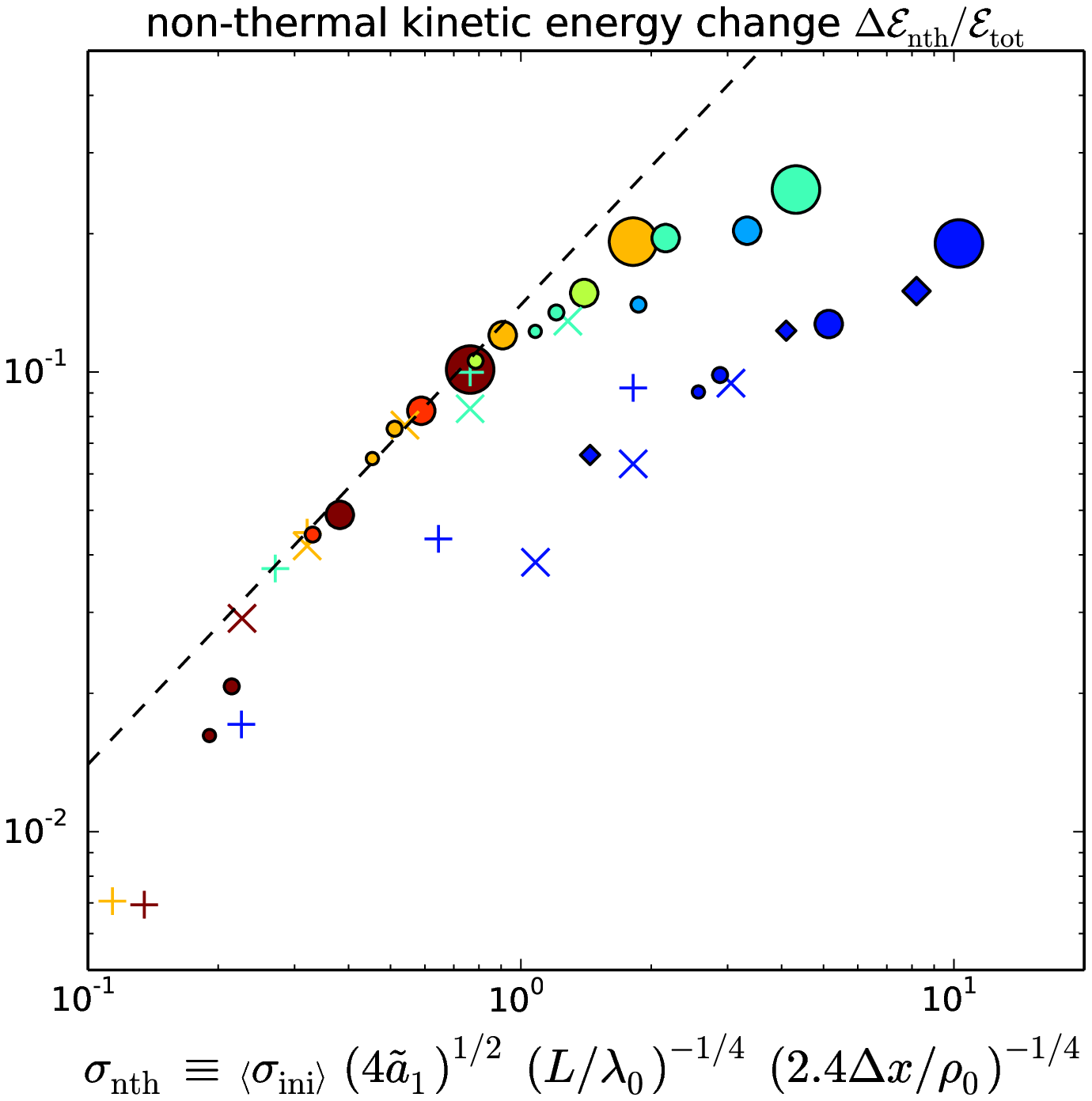}
\caption{Global gain of the particle energy divided into 
thermal $\Delta\mathcal{E}_{\rm th}$ (\emph{left panel})
and non-thermal $\Delta\mathcal{E}_{\rm nth}$ (\emph{right panel})
components, normalised to the total energy $\mathcal{E}_{\rm tot}$,
as functions of modified magnetisation parameters $\sigma_{\rm th}$ and $\sigma_{\rm nth}$, respectively, chosen to minimise scatter around suggested trends (dashed black lines) $\propto\sigma_{\rm th}^{1/3}$ and $\propto\sigma_{\rm nth}$, respectively.
The symbol types are the same as in the right panel of Figure \ref{fig_evo_ele}.}
\label{fig_deltaEk}
\end{figure}

We use the final non-thermal energy fractions $f_E$ to divide the global energy gain of the particles into the non-thermal and thermal parts:
\bea
\Delta\mathcal{E}_{\rm nth} &=& f_E\,\mathcal{E}_{\rm kin,fin}\,,
\\
\Delta\mathcal{E}_{\rm th} &=& (1-f_E)\mathcal{E}_{\rm kin,fin} - \mathcal{E}_{\rm kin,ini}\,,
\eea
%
%
where $\mathcal{E}_{\rm kin,ini} = \mathcal{E}_{\rm tot} - \mathcal{E}_{\rm B,ini}$ and $\mathcal{E}_{\rm kin,fin} \simeq \mathcal{E}_{\rm tot} - \mathcal{E}_{\rm B,fin}$, since by $ct = 25L$ the total electric energy that mediates the dissipation of magnetic energy decreases to the level of $\mathcal{E}_{\rm E,fin} < 10^{-2}\mathcal{E}_{\rm tot}$.
The two components of particle energy gain are presented in Figure \ref{fig_deltaEk} as functions of yet two other modified magnetisation parameters
$\sigma_{\rm th} \equiv \left<\sigma_{\rm ini}\right> (4\tilde{a}_1)^{-1/2} (L/\lambda_0)^{3/4} (2.4\Delta x/\rho_0)^{3/4}$
and $\sigma_{\rm nth} \equiv \left<\sigma_{\rm ini}\right> (4\tilde{a}_1)^{1/2} (L/\lambda_0)^{-1/4} (2.4\Delta x/\rho_0)^{-1/4}$, respectively.
We find that the cases of {\tt para\_k1} (deep blue symbols) stand out from other cases, having significantly lower thermal energy gains, suggesting that they are limited by the magnetic topology.
On the other hand, their non-thermal energy gains are comparable to other cases, but achieved at significantly higher values of $\sigma_{\rm nth}$.
Power-law trends can be suggested only for sufficiently high wavenumbers ($L/\lambda_0 \gtrsim 4\sqrt{2}$): $\Delta\mathcal{E}_{\rm th} \propto \sigma_{\rm th}^{1/3}$ and $\Delta\mathcal{E}_{\rm nth} \propto \sigma_{\rm nth}$, respectively.
However, in the {\tt diag\_k8} cases (brown symbols), a steeper trend for the non-thermal energy gain $\Delta\mathcal{E}_{\rm nth} \propto \sigma_{\rm nth}^{5/2}$ is apparent for low magnetisation values $\sigma_{\rm nth} < 0.25$.
The highest value of $\Delta\mathcal{E}_{\rm nth} / \mathcal{E}_{\rm tot} \simeq 25\%$ is obtained for our large simulation {\tt diag\_k2}.

\section{Discussion}
\label{sec:disc}

Our new results extend the previous study of 2D PIC simulations of ABC fields for the {\tt para\_k1} case in the non-radiative regime \citep{0004-637X-826-2-115}, and connect it with a study of 3D PIC simulations for the cases {\tt diag\_k2} and {\tt diag\_k4} \citep{10.1093/mnras/sty2549}.
They can also be compared with the FF simulations of ABC fields presented in \cite{Zrake_East_2016}.
In particular, the magnetic dissipation efficiency in the FF limit in 2D has been estimated at $\epsilon_{\rm diss} \simeq 70\%$, while our results suggest $\epsilon_{\rm diss} \simeq 62\%$ in the limit of $L/\lambda_0 \gg 1$.
It should be noted, however, that in PIC simulations this limit forces us towards lower magnetisation values.

In \cite{0004-637X-826-2-115}, a relation between the electric energy growth time scale $\tau_{\rm E,peak}$ and the initial characteristic hot magnetisation $\sigma_{\rm hot}$ was suggested in the following form:
\be
\tau_{\rm E,peak} \simeq \frac{0.13}{v_{\rm A}(0.21\sigma_{\rm hot})} \,,
\label{eq_tauE_KN16}
\ee
where $v_{\rm A}(\sigma) \equiv [\sigma/(1+\sigma)]^{1/2}$ was treated as a function in the form of Alfven velocity of arbitrarily scaled argument $\sigma$, and $\sigma_{\rm hot} \equiv \left<\sigma_{\rm ini}\right>/2$ was a characteristic value of hot magnetisation based on $B_0^2$ instead of the mean value $\left<B^2\right>$ used here\footnote{We note that the characteristic values of $\sigma_{\rm hot}$ reported in \cite{0004-637X-826-2-115} were underestimated by a constant factor of $\simeq 1.13$.}.
The above relation is shown in the right panel of Figure \ref{fig_evo_ele} with a dashed blue line (cf. Figure 3 of \citealt{0004-637X-826-2-115}).
We can see that the previously suggested trend agrees very well with the previous measurements from \cite{0004-637X-826-2-115}, and is very close to the new trend line in the range of $1.5 < \left<\sigma_{\rm ini}\right> < 12.5$.
However, the previous trend predicts significantly shorter growth time scales for low magnetisation values $\left<\sigma_{\rm ini}\right> < 1$ that is probed here with simulations for $L/\lambda_0 \ge 4\sqrt{2}$.

Our new scaling described by Eq. (\ref{eq_tauE_fit}) is more natural, without arbitrary scaling parameters.
It suggests that in the FF limit, when $\left<\sigma_{\rm ini}\right> \to \infty$ and $\beta_{\rm A,ini} \to 1$, we should expect that the growth time scale should become $\tau_{\rm E,FF} \simeq 0.233/(L/\lambda_0)$.
For $L/\lambda_0 = \sqrt{2}$, this would yield $\tau_{\rm E,FF} \simeq 0.16$, somewhat longer than $\tau_{\rm E,FF} \simeq 0.13$ indicated by \cite{0004-637X-826-2-115}.
As for why should $\tau_{\rm E,peak} (L/\lambda_0)$ scale with $\beta_{\rm A,ini}^{-3}$ requires a theoretical investigation of the linear coalescence instability beyond the FF limit, with proper treatment of magnetic nulls, which is beyond the scope of this work.

We can only partially confirm a relation between non-thermal energy fraction and initial mean hot magnetisation $f_E \propto \left<\sigma_{\rm ini}\right>^{3/4}$ originally suggested in \cite{0004-637X-826-2-115}.
This relation appears to hold for the {\tt para\_k1} case, including new simulations extending into the $\left<\sigma_{\rm ini}\right> \sim 1$ regime, and possibly also for higher values of $L/\lambda_0$ as long as $\sigma_f > 1$
(see the right panel of Figure \ref{fig_spec_stat}).
However, for the cases where a power-law index $p$ can be determined, a simple linear relation holds between $p$ and the initial magnetic energy fraction $\mathcal{E}_{\rm B,ini} / \mathcal{E}_{\rm tot}$ (see Eq. \ref{eq_p_fit}), at least over the studied range of $0.3 < \mathcal{E}_{\rm B,ini} / \mathcal{E}_{\rm tot} < 0.9$ (see the left panel of Figure \ref{fig_spec_stat}).

We have introduced several \emph{modified magnetisation} parameters, as combinations of the initial mean magnetisation $\left<\sigma_{\rm ini}\right>$ with other input parameters, in order to describe the scalings of global output parameters.
The particular formulae for the modified magnetisations were chosen in order to minimise scatter around the suggested trends, with the exponents of $\Delta x/\rho_0$, $L/\rho_0$, $L/\lambda_0$ and $\tilde{a}_1$ estimated empirically with the accuracy of $\sim \pm 1/4$.
The energy conservation accuracy for ABC fields simulated with the {\tt Zeltron} code is found to scale roughly like $\delta_\mathcal{E} \propto \left<\sigma_{\rm ini}\right>^{-5/2} (\Delta x/\rho_0)^2 (L/\rho_0)^2$, not sensitive to $\lambda_0$.
This is different from the reference case of uniform magnetic field, in which we found $\delta_\mathcal{E} \propto \left<\sigma_{\rm ini}\right>^{-1} (\Delta x/\rho_0)^2$, independent of $L$.
On the other hand, the magnetic helicity conservation accuracy is found to scale like $\delta_\mathcal{H} \propto (\lambda_0/\rho_0)^{-2}$, but it is not sensitive to $\Delta x/\rho_0$ or $L/\rho_0$.
This is in contrast to the force-free simulations of \cite{Zrake_East_2016}, in which $\delta_\mathcal{H} \propto (\Delta x)^{2.8}$.
Further investigation is required in order to explain these differences.

For the non-thermal energy fraction $f_{\rm E}$, the scaling with initial mean magnetisation $\left<\sigma_{\rm ini}\right>$ is rather ambiguous.
Only in the special case of $L/\lambda_0 = \sqrt{2}$ we have sufficient range of $\left<\sigma_{\rm ini}\right>$ values to claim that $f_{\rm E} \propto \left<\sigma_{\rm ini}\right>^{3/4}$; this scaling is improved by additional dependence on the particle anisotropy level $\tilde{a}_1$.
The scalings of thermal and non-thermal kinetic energy gains, $\Delta\mathcal{E}_{\rm th}$ and $\Delta\mathcal{E}_{\rm nth}$, respectively, can in principle be derived from the scalings of $f_{\rm E}$ and magnetic dissipation efficiency $\epsilon_{\rm diss}$.
The ambiguity of the $f_{\rm E}$ scaling makes it not straightforward to predict in detail the scalings of $\Delta\mathcal{E}_{\rm th}$ and $\Delta\mathcal{E}_{\rm nth}$.

The initial mean hot magnetisation $\left<\sigma_{\rm ini}\right>$ of ABC fields with relativistically warm plasma ($\Theta = 1$) is strongly limited by the simulation size, especially if one would like to resolve numerically all the fundamental length scales, in particular the nominal gyroradius $\rho_0$.
For a given effective wavenumber $L/\lambda_0$, higher values of $\left<\sigma_{\rm ini}\right>$ can only be reached by increasing the system size $L/\rho_0$
\footnote{One can achieve a somewhat higher $\left<\sigma_{\rm ini}\right>$ by increasing the local particle anisotropic parameter $\tilde{a}_1$. However, some numerical artefacts are observed for $\tilde{a}_1 \simeq 1/2$.}.
It can be expected that larger simulations would show more effective non-thermal particle acceleration with harder high-energy tails indicated by higher values of non-thermal energy fractions $f_e$ and lower values of power-law indices $p$.
Eventually, at sufficiently high $\left<\sigma_{\rm ini}\right>$, and with $L/\lambda_0 \ge 2$, it should be possible to achieve particle distributions dominated energetically by the high-energy particles, with $p < 2$, as has been demonstrated in the case of Harris-layer reconnection \citep{Sironi_2014,PhysRevLett.113.155005,Werner_2016,10.1093/mnras/sty452}.
What remains unclear, though, is the level of thermal energy gains.

Our results show that the case {para\_k1} characterised by the lowest unstable effective wavenumber $L/\lambda_0 = \sqrt{2}$, studied in detail by \cite{0004-637X-826-2-115} and \cite{Yuan_2016}, has a limited efficiency of both thermal and non-thermal particle acceleration, which is related to the limited magnetic dissipation efficiency.
On the other hand, 2D ABC fields with high $L/\lambda_0$ values, although also limited by topological constraints \citep{Zrake_East_2016}, can be used as a model for kinetic investigations of decaying relativistic magnetised turbulence, an alternative to uncorrelated magnetic fluctuations \citep{PhysRevLett.121.255101,Comisso_2019,Comisso_2020}.
Relativistic magnetised turbulence has also been investigated extensively by means of PIC simulations in the driven mode \citep{PhysRevLett.118.055103,10.1093/mnras/stx2883,Zhdankin_2018,PhysRevLett.122.055101,10.1093/mnras/staa284,Wong_2020}.

\vskip 1em
These results are based on numerical simulations performed at the supercomputer {\tt Prometheus} located at the Academic Computer Centre `Cyfronet' of the AGH University of Science and Technology in Krakow, Poland (PLGrid grants {\tt plgpic20,ehtsim});
and at the computing cluster {\tt Chuck} located at the Nicolaus Copernicus Astronomical Center of the Polish Academy of Sciences in Warsaw, Poland.
QC and KN were supported by the Polish National Science Center grant 2015/18/E/ST9/00580.
BM acknowledges support from DOE through the LDRD program at LANL and NASA Astrophysics Theory Program.

\bibliographystyle{jpp}

\begin{thebibliography}{75}
\expandafter\ifx\csname natexlab\endcsname\relax\def\natexlab#1{#1}\fi
\def\au#1{#1} \def\ed#1{#1} \def\yr#1{#1}\def\at#1{#1}\def\jt#1{\textit{#1}}
  \def\bt#1{#1}\def\bvol#1{\textbf{#1}} \def\vol#1{#1} \def\pg#1{#1}
  \def\publ#1{#1}\def\arxiv#1{#1}\def\org#1{#1}\def\st#1{\textit{#1}}

\bibitem[Abdo {\em et~al.\/}(2011{\natexlab{{\em a\/}}})]{Abdo739}
{\sc \au{Abdo, A.~A.}, et al.} \yr{2011{\natexlab{{\em
  a\/}}}}  \at{Gamma-ray flares from the Crab Nebula}.  \jt{Science}
  \bvol{331}~(6018),  \pg{739--742}.

\bibitem[Abdo {\em et~al.\/}(2011{\natexlab{{\em b\/}}})]{Abdo_2011}
{\sc \au{Abdo, A.~A.}, et al.}
  \yr{2011{\natexlab{{\em b\/}}}}  \at{Fermi Gamma-ray Space Telescope observations
of the gamma-ray outburst from 3C454.3 in November 2010}.  \jt{Astrophys. J.}  \bvol{733}~(2),
  \pg{L26}.

\bibitem[Ackermann {\em et~al.\/}(2016)]{Ackermann_2016}
{\sc \au{Ackermann, M.}, et al.}
  \yr{2016}  \at{{Minute}-{timescale} $>$ 100 {MeV} $\gamma$ -{ray} {variability}
  {during} {the} {giant} {outburst} {of} {quasar} 3C 279 {observed} {by}
  {Fermi}-{LAT} {in} 2015 {June}}.  \jt{Astrophys. J.}
  \bvol{824}~(2),  \pg{L20}.

\bibitem[Aharonian {\em et~al.\/}(2007)]{Aharonian_2007}
{\sc \au{Aharonian, F.}, et al.} \yr{2007}  \at{An exceptional very high energy
  gamma-ray flare of {PKS} 2155-304}.  \jt{Astrophys. J.}
  \bvol{664}~(2),  \pg{L71--L74}.

\bibitem[Albert {\em et~al.\/}(2007)]{Albert_2007}
{\sc \au{Albert, J.}, et al.} \yr{2007}
  \at{Variable very high energy gamma-ray emission from Markarian 501}.
  \jt{Astrophys. J.}  \bvol{669}~(2),  \pg{862--883}.

\bibitem[Aleksi{\'{c}} {\em et~al.\/}(2011)]{Aleksi__2011}
{\sc \au{Aleksi{\'{c}}, J.}, et al.} \yr{2011}  \at{{Magic} {discovery} {of} {very} {high}
  {energy} {emission} {from} {the} {FSRQ} {PKS} 1222+21},  \jt{Astrophys. J.}  \bvol{730}~(1),  \pg{L8}.


\bibitem[Arnold(1965)]{Arnold_1965}
{\sc \au{Arnold, V.}} \yr{1965}  \at{Sur une proprietes topologique des applications globalment canonique de la mechanique classique},  \jt{CR. Acad. Sci. Paris}  \bvol{261},  \pg{3719-3722}.

\bibitem[Arons(2012)]{Arons_2012}
{\sc \au{Arons, J.}} \yr{2012}  \at{Pulsar wind nebulae as cosmic pevatrons: A
  current sheet’s tale}.  \jt{Space Sci. Rev.}  \bvol{173}~(1-4),
  \pg{341–367}.

\bibitem[Bessho \& Bhattacharjee(2012)]{Bessho_2012}
{\sc \au{Bessho, N.} \& \au{Bhattacharjee, A.}} \yr{2012}  \at{{Fast}
 {magnetic} {reconnection} {and} {particle} {acceleration} {in} {relativistic}
  {low}-{density} {electron}-{positron} {plasmas} {without} {guide} {field}}.
  \jt{Astrophys. J.}  \bvol{750}~(2),  \pg{129}.

\bibitem[Blandford {\em et~al.\/}(2017)Blandford, Yuan, Hoshino \&
  Sironi]{Blandford_2017}
{\sc \au{Blandford, R.}, \au{Yuan, Y.}, \au{Hoshino, M.} \& \au{Sironi, L.}}
  \yr{2017}  \at{Magnetoluminescence}.  \jt{Space Sci. Rev.}
  \bvol{207}~(1-4),  \pg{291–317}.

\bibitem[Buehler {\em et~al.\/}(2012)Buehler, Scargle, Blandford, Baldini,
  Baring, Belfiore, Charles, Chiang, D'Ammando, Dermer, Funk, Grove, Harding,
  Hays, Kerr, Massaro, Mazziotta, Romani, Parkinson, Tennant \&
  Weisskopf]{Buehler_2012}
{\sc \au{Buehler, R.}, et al.} \yr{2012}  \at{{Gamma}-{ray}
  {activity} {in} {the} {Crab} {Nebula}: {the} {exceptional} {flare} {of} 2011
  {April}}.  \jt{Astrophys. J.}  \bvol{749}~(1),  \pg{26}.

\bibitem[Buehler \& Blandford(2014)]{B_hler_2014}
{\sc \au{Buehler, R.} \& \au{Blandford, R.}} \yr{2014}  \at{The surprising Crab
  pulsar and its nebula: a review}.  \jt{Rep. Prog. Phys.}
  \bvol{77}~(6),  \pg{066901}.

\bibitem[Cerutti {\em et~al.\/}(2012)Cerutti, Werner, Uzdensky \&
  Begelman]{Cerutti_2012}
{\sc \au{Cerutti, B.}, \au{Werner, G.~R.}, \au{Uzdensky, D.~A.} \&
  \au{Begelman, M.~C.}} \yr{2012}  \at{{Beaming} {and} {rapid} {variability}
  {of} {high}-{energy} {radiation} {from} {relativistic} {pair} {plasma}
  {reconnection}}.  \jt{Astrophys. J.}  \bvol{754}~(2),  \pg{L33}.

\bibitem[Cerutti {\em et~al.\/}(2013)Cerutti, Werner, Uzdensky \&
  Begelman]{0004-637X-770-2-147}
{\sc \au{Cerutti, B.}, \au{Werner, G.~R.}, \au{Uzdensky, D.~A.} \&
  \au{Begelman, M.~C.}} \yr{2013}  \at{Simulations of particle acceleration
  beyond the classical synchrotron burnoff limit in magnetic reconnection: An
  explanation of the Crab flares}.  \jt{Astrophys. J.}
  \bvol{770}~(2),  \pg{147}.

\bibitem[Cerutti {\em et~al.\/}(2014)Cerutti, Werner, Uzdensky \&
  Begelman]{Cerutti_2014}
{\sc \au{Cerutti, B.}, \au{Werner, G.~R.}, \au{Uzdensky, D.~A.} \&
  \au{Begelman, M.~C.}} \yr{2014}  \at{{Three}-{dimensional} {relativistic}
  {pair} {plasma} {reconnection} {with} {radiative} {feedback} {in} {the}
  {Crab} {Nebula}}.  \jt{Astrophys. J.}  \bvol{782}~(2),  \pg{104}.

\bibitem[Christie {\em et~al.\/}(2018)Christie, Petropoulou, Sironi \&
  Giannios]{10.1093/mnras/sty2636}
{\sc \au{Christie, I.~M.}, \au{Petropoulou, M.}, \au{Sironi, L.} \&
  \au{Giannios, D.}} \yr{2018}  \at{{Radiative signatures of plasmoid-dominated
  reconnection in blazar jets}}.  \jt{Mon. Not. R. Astron. Soc.}  \bvol{482}~(1),  \pg{65--82}.

\bibitem[Clausen-Brown \& Lyutikov(2012)]{10.1111/j.1365-2966.2012.21349.x}
{\sc \au{Clausen-Brown, E.} \& \au{Lyutikov, M.}} \yr{2012}  \at{{Crab Nebula
  gamma-ray flares as relativistic reconnection minijets}}.  \jt{Mon. Not. R. Astron. Soc.}  \bvol{426}~(2),  \pg{1374--1384}.

\bibitem[Comisso \& Sironi(2018)]{PhysRevLett.121.255101}
{\sc \au{Comisso, L.} \& \au{Sironi, L.}} \yr{2018}  \at{Particle acceleration
  in relativistic plasma turbulence}.  \jt{Phys. Rev. Lett.}  \bvol{121},
  \pg{255101}.

\bibitem[Comisso \& Sironi(2019)]{Comisso_2019}
{\sc \au{Comisso, L.} \& \au{Sironi, L.}} \yr{2019}  \at{The interplay of
  magnetically dominated turbulence and magnetic reconnection in producing
  nonthermal particles}.  \jt{Astrophys. J.}  \bvol{886}~(2),
  \pg{122}.

\bibitem[Comisso {\em et~al.\/}(2020)Comisso, Sobacchi \& Sironi]{Comisso_2020}
{\sc \au{Comisso, L.}, \au{Sobacchi, E.} \& \au{Sironi, L.}} \yr{2020}
  \at{Hard synchrotron spectra from magnetically dominated plasma turbulence}.
  \jt{Astrophys. J.}  \bvol{895}~(2),  \pg{L40}.

\bibitem[East {\em et~al.\/}(2015)East, Zrake, Yuan \&
  Blandford]{PhysRevLett.115.095002}
{\sc \au{East, W.~E.}, \au{Zrake, J.}, \au{Yuan, Y.} \& \au{Blandford, R.~D.}}
  \yr{2015}  \at{Spontaneous decay of periodic magnetostatic equilibria}.
  \jt{Phys. Rev. Lett.}  \bvol{115},  \pg{095002}.

\bibitem[Giannios(2013)]{10.1093/mnras/stt167}
{\sc \au{Giannios, D.}} \yr{2013}  \at{{Reconnection-driven plasmoids in
  blazars: fast flares on a slow envelope}}.  \jt{Mon. Not. R. Astron. Soc.}  \bvol{431}~(1),  \pg{355--363}.

\bibitem[Giannios {\em et~al.\/}(2009)Giannios, Uzdensky \&
  Begelman]{10.1111/j.1745-3933.2009.00635.x}
{\sc \au{Giannios, D.}, \au{Uzdensky, D.~A.} \& \au{Begelman, M.~C.}} \yr{2009}
   \at{{Fast TeV variability in blazars: jets in a jet}}.  \jt{Mon. Not. R. Astron. Soc.} \bvol{395}~(1),  \pg{L29--L33}.

\bibitem[Guo {\em et~al.\/}(2014)Guo, Li, Daughton \&
  Liu]{PhysRevLett.113.155005}
{\sc \au{Guo, F.}, \au{Li, H.}, \au{Daughton, W.} \& \au{Liu, Y.-H.}} \yr{2014}
   \at{Formation of hard power laws in the energetic particle spectra resulting
  from relativistic magnetic reconnection}.  \jt{Phys. Rev. Lett.}  \bvol{113},
   \pg{155005}.

\bibitem[Guo {\em et~al.\/}(2019)Guo, Li, Daughton, Kilian, Li, Liu, Yan \&
  Ma]{Guo_2019}
{\sc \au{Guo, F.}, \au{Li, X.}, \au{Daughton, W.}, \au{Kilian, P.}, \au{Li,
  H.}, \au{Liu, Y.-H.}, \au{Yan, W.} \& \au{Ma, D.}} \yr{2019}  \at{Determining
  the dominant acceleration mechanism during relativistic magnetic reconnection
  in large-scale systems}.  \jt{Astrophys. J.}  \bvol{879}~(2),
  \pg{L23}.

\bibitem[Guo {\em et~al.\/}(2020)Guo, Li, Daughton, Li, Kilian, Liu, Zhang \&
  Zhang]{guo2020magnetic}
{\sc \au{Guo, F.}, \au{Li, X.}, \au{Daughton, W.}, \au{Li, H.}, \au{Kilian,
  P.}, \au{Liu, Y.-H.}, \au{Zhang, Q.} \& \au{Zhang, H.}} \yr{2020} Magnetic
  energy release, plasma dynamics and particle acceleration during relativistic
  turbulent magnetic reconnection,  \arxiv{arXiv: 2008.02743}.

\bibitem[Guo {\em et~al.\/}(2016)Guo, Li, Li, Daughton, Zhang, Lloyd-Ronning,
  Liu, Zhang \& Deng]{Guo_2016}
{\sc \au{Guo, F.}, \au{Li, X.}, \au{Li, H.}, \au{Daughton, W.}, \au{Zhang, B.},
  \au{Lloyd-Ronning, N.}, \au{Liu, Y.-H.}, \au{Zhang, H.} \& \au{Deng, W.}}
  \yr{2016}  \at{{Efficient} {production} {of} {high}-{energy} {nonthermal}
  {particles} {during} {magnetic} {reconnection} {in} a {magnetically}
  {dominated} {ion}{\textendash}{electron} {plasma}}.  \jt{Astrophys. J.}  \bvol{818}~(1),  \pg{L9}.

\bibitem[Guo {\em et~al.\/}(2015)Guo, Liu, Daughton \& Li]{Guo_2015}
{\sc \au{Guo, F.}, \au{Liu, Y.-H.}, \au{Daughton, W.} \& \au{Li, H.}} \yr{2015}
   \at{{Particle} {acceleration} {and} {plasma} {dynamics} {during} {magnetic}
  {reconnection} {in} {the} {magnetically} {dominated} {regime}}.  \jt{Astrophys. J.}  \bvol{806}~(2),  \pg{167}.

\bibitem[Hoshino (2012)]{2012PhRvL.108m5003H}
{\sc \au{Hoshino, Masahiro}} \yr{2012}  \at{Stochastic Particle Acceleration in Multiple Magnetic Islands during Reconnection}.  \jt{Phys. Rev. Lett.}  \bvol{108},
  \pg{135003}.

\bibitem[Jaroschek {\em et~al.\/}(2004)Jaroschek, Treumann, Lesch \&
  Scholer]{doi:10.1063/1.1644814}
{\sc \au{Jaroschek, C.~H.}, \au{Treumann, R.~A.}, \au{Lesch, H.} \&
  \au{Scholer, M.}} \yr{2004}
  \at{Fast reconnection in relativistic pair plasmas: analysis of particle acceleration in self-consistent full particle simulations}.
  \jt{Phys. Plasmas}  \bvol{11}~(3),  \pg{1151--1163}.

\bibitem[Kagan {\em et~al.\/}(2013)Kagan, Milosavljevi{\'{c}} \&
  Spitkovsky]{Kagan_2013}
{\sc \au{Kagan, D.}, \au{Milosavljevi{\'{c}}, M.} \& \au{Spitkovsky, A.}}
  \yr{2013}  \at{A {flux} {rope} {network} {and} {particle} {acceleration} {in}
  {three}-{dimensional} {relativistic} {magnetic} {reconnection}}.  \jt{Astrophys. J.}  \bvol{774}~(1),  \pg{41}.

\bibitem[Kagan {\em et~al.\/}(2016)Kagan, Nakar \& Piran]{Kagan_2016}
{\sc \au{Kagan, D.}, \au{Nakar, E.} \& \au{Piran, T.}} \yr{2016}  \at{{Beaming}
  {of} {particles} {and} {synchrotron} {radiation} {in} {relativistic}
  {magnetic} {reconnection}}.  \jt{Astrophys. J.}  \bvol{826}~(2),
  \pg{221}.

\bibitem[Kagan {\em et~al.\/}(2018)Kagan, Nakar \& Piran]{10.1093/mnras/sty452}
{\sc \au{Kagan, D.}, \au{Nakar, E.} \& \au{Piran, T.}} \yr{2018}  \at{{Physics
  of the saturation of particle acceleration in relativistic magnetic
  reconnection}}.  \jt{Mon. Not. R. Astron. Soc.}
  \bvol{476}~(3),  \pg{3902--3912}.

\bibitem[Kirk \& Skjaraasen(2003)]{Kirk_2003}
{\sc \au{Kirk, J.~G.} \& \au{Skjaraasen, O.}} \yr{2003}  \at{Dissipation in
  poynting-flux{\textendash}dominated flows: The sigma-problem of the Crab
  pulsar wind}.  \jt{Astrophys. J.}  \bvol{591}~(1),
  \pg{366--379}.

\bibitem[Komissarov(2012)]{10.1093/mnras/sts214}
{\sc \au{Komissarov, S.~S.}} \yr{2012}  \at{{Magnetic dissipation in the Crab
  Nebula}}.  \jt{Mon. Not. R. Astron. Soc.}
  \bvol{428}~(3),  \pg{2459--2466}.

\bibitem[Komissarov \& Lyutikov(2011)]{10.1111/j.1365-2966.2011.18516.x}
{\sc \au{Komissarov, S.~S.} \& \au{Lyutikov, M.}} \yr{2011}  \at{{On the origin
  of variable gamma-ray emission from the Crab Nebula}}.  \jt{Mon. Not. R. Astron. Soc.}  \bvol{414}~(3),  \pg{2017--2028}.

\bibitem[Liu {\em et~al.\/}(2011)Liu, Li, Yin, Albright, Bowers \&
  Liang]{doi:10.1063/1.3589304}
{\sc \au{Liu, W.}, \au{Li, H.}, \au{Yin, L.}, \au{Albright, B.~J.}, \au{Bowers,
  K.~J.} \& \au{Liang, E.~P.}} \yr{2011}  \at{Particle energization in 3D
  magnetic reconnection of relativistic pair plasmas}.  \jt{Phys. Plasmas}
   \bvol{18}~(5),  \pg{052105}.

\bibitem[Lyubarsky \& Liverts(2008)]{Lyubarsky_2008}
{\sc \au{Lyubarsky, Y.} \& \au{Liverts, M.}} \yr{2008}  \at{Particle
  acceleration in the driven relativistic reconnection}.  \jt{Astrophys. J.}  \bvol{682}~(2),  \pg{1436--1442}.

\bibitem[Lyubarsky(2012)]{10.1111/j.1365-2966.2012.22097.x}
{\sc \au{Lyubarsky, Y.~E.}} \yr{2012}  \at{{Highly magnetized region in pulsar
  wind nebulae and origin of the Crab gamma-ray flares}}.  \jt{Mon. Not. R. Astron. Soc.}  \bvol{427}~(2),  \pg{1497--1502}.

\bibitem[Lyutikov {\em et~al.\/}(2018)Lyutikov, Komissarov, Sironi \&
  Porth]{lyutikov_komissarov_sironi_porth_2018}
{\sc \au{Lyutikov, M.}, \au{Komissarov, S.}, \au{Sironi, L.} \& \au{Porth, O.}}
  \yr{2018}  \at{Particle acceleration in explosive relativistic reconnection
  events and Crab Nebula gamma-ray flares}.  \jt{J. Plasma Phys.}
  \bvol{84}~(2),  \pg{635840201}.

\bibitem[Lyutikov {\em et~al.\/}(2017{\natexlab{{\em a\/}}})Lyutikov, Sironi,
  Komissarov \& Porth]{lyutikov_sironi_komissarov_porth_2017}
{\sc \au{Lyutikov, M.}, \au{Sironi, L.}, \au{Komissarov, S.~S.} \& \au{Porth,
  O.}} \yr{2017{\natexlab{{\em a\/}}}}  \at{Explosive x-point collapse in
  relativistic magnetically dominated plasma}.  \jt{J. Plasma Phys.}
  \bvol{83}~(6),  \pg{635830601}.

\bibitem[Lyutikov {\em et~al.\/}(2017{\natexlab{{\em b\/}}})Lyutikov, Sironi,
  Komissarov \& Porth]{Lyutikov2017}
{\sc \au{Lyutikov, M.}, \au{Sironi, L.}, \au{Komissarov, S.~S.} \& \au{Porth,
  O.}} \yr{2017{\natexlab{{\em b\/}}}}  \at{Particle acceleration in
  relativistic magnetic flux-merging events}.  \jt{J. Plasma Phys.}
  \bvol{83}~(6),  \pg{635830602}.

\bibitem[Mayer {\em et~al.\/}(2013)Mayer, Buehler, Hays, Cheung, Dutka, Grove,
  Kerr \& Ojha]{Mayer_2013}
{\sc \au{Mayer, M.}, \au{Buehler, R.}, \au{Hays, E.}, \au{Cheung, C.~C.},
  \au{Dutka, M.~S.}, \au{Grove, J.~E.}, \au{Kerr, M.} \& \au{Ojha, R.}}
  \yr{2013}  \at{{Rapid} {gamma}-{ray} {flux} {variability} {during} {the} 2013
  {March} {Crab} {Nebula} {flare}}.  \jt{Astrophys. J.}
  \bvol{775}~(2),  \pg{L37}.

\bibitem[Mehlhaff {\em et~al.\/}(2020)Mehlhaff, Werner, Uzdensky \&
  Begelman]{10.1093/mnras/staa2346}
{\sc \au{Mehlhaff, J.~M.}, \au{Werner, G.~R.}, \au{Uzdensky, D.~A.} \&
  \au{Begelman, M.~C.}} \yr{2020}  \at{{Kinetic beaming in radiative
  relativistic magnetic reconnection: a mechanism for rapid gamma-ray flares in
  jets}}.  \jt{Mon. Not. R. Astron. Soc.}
  \bvol{498}~(1),  \pg{799--820}.

\bibitem[{Melzani, Micka\"el} {\em et~al.\/}(2014){Melzani, Micka\"el},
  {Walder, Rolf}, {Folini, Doris}, {Winisdoerffer, Christophe} \& {Favre, Jean
  M.}]{refId0}
{\sc \au{{Melzani, Micka\"el}}, \au{{Walder, Rolf}}, \au{{Folini, Doris}},
  \au{{Winisdoerffer, Christophe}} \& \au{{Favre, Jean M.}}} \yr{2014}  \at{The
  energetics of relativistic magnetic reconnection: ion-electron repartition
  and particle distribution hardness}.  \jt{A\&A}  \bvol{570},  \pg{A112}.

\bibitem[Nalewajko(2013)]{10.1093/mnras/sts711}
{\sc \au{Nalewajko, K.}} \yr{2013}  \at{{The brightest gamma-ray flares of
  blazars}}.  \jt{Mon. Not. R. Astron. Soc.}
  \bvol{430}~(2),  \pg{1324--1333}.

\bibitem[Nalewajko(2018)]{nalewajko2018relativistic}
{\sc \au{Nalewajko, K.}} \yr{2018} Relativistic magnetic reconnection in
  application to gamma-ray astrophysics.  \bt{In {\em XXXVIII Polish
  Astronomical Society Meeting\/} (ed. \ed{Agata Rozanska})},
  \vol{vol.~7},  \pg{pp. 310--315},  \arxiv{arXiv: 1808.00478}.

\bibitem[{Nalewajko}(2018)]{10.1093/mnras/sty2549}
{\sc \au{{Nalewajko}, K.}} \yr{2018}  \at{{Three-dimensional kinetic
  simulations of relativistic magnetostatic equilibria}}.  \jt{Mon. Not. R. Astron. Soc.}
  \bvol{481},  \pg{4342--4354}.

\bibitem[Nalewajko {\em et~al.\/}(2012)Nalewajko, Begelman, Cerutti, Uzdensky
  \& Sikora]{10.1111/j.1365-2966.2012.21721.x}
{\sc \au{Nalewajko, K.}, \au{Begelman, M.~C.}, \au{Cerutti, B.}, \au{Uzdensky,
  D.~A.} \& \au{Sikora, M.}} \yr{2012}  \at{{Energetic constraints on a rapid
  gamma-ray flare in PKS 1222+216}}.  \jt{Mon. Not. R. Astron. Soc.}  \bvol{425}~(4),  \pg{2519--2529}.

\bibitem[Nalewajko {\em et~al.\/}(2011)Nalewajko, Giannios, Begelman, Uzdensky
  \& Sikora]{10.1111/j.1365-2966.2010.18140.x}
{\sc \au{Nalewajko, K.}, \au{Giannios, D.}, \au{Begelman, M.~C.}, \au{Uzdensky,
  D.~A.} \& \au{Sikora, M.}} \yr{2011}  \at{{Radiative properties of
  reconnection-powered minijets in blazars}}.  \jt{Mon. Not. R. Astron. Soc.}  \bvol{413}~(1),  \pg{333--346}.

\bibitem[Nalewajko {\em et~al.\/}(2018)Nalewajko, Yuan \&
  Chru{\'s}li{\'n}ska]{nalewajko_yuan_chruslinska_2018}
{\sc \au{Nalewajko, K.}, \au{Yuan, Y.} \& \au{Chru{\'s}li{\'n}ska, M.}} \yr{2018}
  \at{{Kinetic simulations of relativistic magnetic reconnection with
  synchrotron and inverse Compton cooling}}.  \jt{J. Plasma Phys.}
  \bvol{84}~(3),  \pg{037501}.

\bibitem[Nalewajko {\em et~al.\/}(2016)Nalewajko, Zrake, Yuan, East \&
  Blandford]{0004-637X-826-2-115}
{\sc \au{Nalewajko, K.}, \au{Zrake, J.}, \au{Yuan, Y.}, \au{East, W.~E.} \&
  \au{Blandford, R.~D.}} \yr{2016}  \at{Kinetic simulations of the lowest-order
  unstable mode of relativistic magnetostatic equilibria}.  \jt{Astrophys. J.}  \bvol{826}~(2),  \pg{115}.

\bibitem[Ortu\~no-Mac\'ias \& Nalewajko(2020)]{10.1093/mnras/staa1899}
{\sc \au{Ortuño-Macías, J.} \& \au{Nalewajko, K.}} \yr{2020}  \at{{Radiative
  kinetic simulations of steady-state relativistic plasmoid magnetic
  reconnection}}.  \jt{Mon. Not. R. Astron. Soc.}
  \bvol{497}~(2),  \pg{1365--1381}.

\bibitem[Petropoulou {\em et~al.\/}(2016)Petropoulou, Giannios \&
  Sironi]{10.1093/mnras/stw1832}
{\sc \au{Petropoulou, M.}, \au{Giannios, D.} \& \au{Sironi, L.}} \yr{2016}
  \at{{Blazar flares powered by plasmoids in relativistic reconnection}}.
  \jt{Mon. Not. R. Astron. Soc.}  \bvol{462}~(3),
  \pg{3325--3343}.

\bibitem[Petropoulou \& Sironi(2018)]{10.1093/mnras/sty2702}
{\sc \au{Petropoulou, M.} \& \au{Sironi, L.}} \yr{2018}  \at{{The steady growth
  of the high-energy spectral cut-off in relativistic magnetic reconnection}}.
  \jt{Mon. Not. R. Astron. Soc.}  \bvol{481}~(4),
  \pg{5687--5701}.

\bibitem[Petropoulou {\em et~al.\/}(2019)Petropoulou, Sironi, Spitkovsky \&
  Giannios]{Petropoulou_2019}
{\sc \au{Petropoulou, M.}, \au{Sironi, L.}, \au{Spitkovsky, A.} \&
  \au{Giannios, D.}} \yr{2019}  \at{Relativistic magnetic reconnection in
  electron{\textendash}positron{\textendash}proton plasmas: Implications for
  jets of active galactic nuclei}.  \jt{Astrophys. J.}
  \bvol{880}~(1),  \pg{37}.

\bibitem[Sironi {\em et~al.\/}(2015)Sironi, Petropoulou \&
  Giannios]{10.1093/mnras/stv641}
{\sc \au{Sironi, L.}, \au{Petropoulou, M.} \& \au{Giannios, D.}} \yr{2015}
  \at{{Relativistic jets shine through shocks or magnetic reconnection?}}
  \jt{Mon. Not. R. Astron. Soc.}  \bvol{450}~(1),
  \pg{183--191}.

\bibitem[Sironi \& Spitkovsky(2014)]{Sironi_2014}
{\sc \au{Sironi, L.} \& \au{Spitkovsky, A.}} \yr{2014}  \at{{Relativistic}
  {reconnection}: {an} {efficient} {source} {of} {non}-{thermal} {particles}}.
  \jt{Astrophys. J.}  \bvol{783}~(1),  \pg{L21}.

\bibitem[Striani {\em et~al.\/}(2013)Striani, Tavani, Vittorini, Donnarumma,
  Giuliani, Pucella, Argan, Bulgarelli, Colafrancesco, Cardillo, Costa, Monte,
  Ferrari, Mereghetti, Pacciani, Pellizzoni, Piano, Pittori, Rapisarda,
  Sabatini, Soffitta, Trifoglio, Trois, Vercellone \& Verrecchia]{Striani_2013}
{\sc \au{Striani, E.}, et al.} \yr{2013}  \at{{Variable}
  {gamma}-{ray} {emission} {from} {the} {Crab} {Nebula}: {short} {flares} {and}
  {long} {\textquotedblleft}{waves}{\textquotedblright}}.  \jt{Astrophys. J.}  \bvol{765}~(1),  \pg{52}.

\bibitem[Tavani {\em et~al.\/}(2011)]{Tavani736}
{\sc \au{Tavani, M.}, et al.} \yr{2011}  \at{Discovery of powerful gamma-ray flares
  from the Crab Nebula}.  \jt{Science}  \bvol{331}~(6018),  \pg{736--739}.

\bibitem[Taylor(1974)]{PhysRevLett.33.1139}
{\sc \au{Taylor, J.~B.}} \yr{1974}  \at{Relaxation of toroidal plasma and
  generation of reverse magnetic fields}.  \jt{Phys. Rev. Lett.}  \bvol{33},
  \pg{1139--1141}.

\bibitem[Uzdensky {\em et~al.\/}(2011)Uzdensky, Cerutti \&
  Begelman]{Uzdensky_2011}
{\sc \au{Uzdensky, D.~A.}, \au{Cerutti, B.} \& \au{Begelman, M.~C.}} \yr{2011}
  \at{{Reconnection}-{powered} {linear} {accelerator} {and} {gamma}-{ray}
  {flares} {in} {the} {Crab} {Nebula}}.  \jt{Astrophys. J.}
  \bvol{737}~(2),  \pg{L40}.

\bibitem[Werner \& Uzdensky(2017)]{Werner_2017}
{\sc \au{Werner, G.~R.} \& \au{Uzdensky, D.~A.}} \yr{2017}  \at{Nonthermal
  particle acceleration in 3D relativistic magnetic reconnection in pair
  plasma}.  \jt{Astrophys. J.}  \bvol{843}~(2),  \pg{L27}.

\bibitem[{Werner} {\em et~al.\/}(2018){Werner}, {Uzdensky}, {Begelman},
  {Cerutti} \& {Nalewajko}]{10.1093/mnras/stx2530}
{\sc \au{{Werner}, G.~R.}, \au{{Uzdensky}, D.~A.}, \au{{Begelman}, M.~C.},
  \au{{Cerutti}, B.} \& \au{{Nalewajko}, K.}} \yr{2018}  \at{{Non-thermal
  particle acceleration in collisionless relativistic electron-proton
  reconnection}}.   \jt{Mon. Not. R. Astron. Soc.}  \bvol{473},  \pg{4840--4861}.

\bibitem[Werner {\em et~al.\/}(2016)Werner, Uzdensky, Cerutti, Nalewajko \&
  Begelman]{Werner_2016}
{\sc \au{Werner, G.~R.}, \au{Uzdensky, D.~A.}, \au{Cerutti, B.}, \au{Nalewajko,
  K.} \& \au{Begelman, M.~C.}} \yr{2016}  \at{{The} {extent} {of} {power}-{law}
  {energy} {spectra} {in} {collisionless} {relativistic} {magnetic}
  {reconnection} {in} {pair} {plasmas}}.  \jt{Astrophys. J.}
  \bvol{816}~(1),  \pg{L8}.

\bibitem[Wong {\em et~al.\/}(2020)Wong, Zhdankin, Uzdensky, Werner \&
  Begelman]{Wong_2020}
{\sc \au{Wong, K.}, \au{Zhdankin, V.}, \au{Uzdensky, D.~A.}, \au{Werner, G.~R.}
  \& \au{Begelman, M.~C.}} \yr{2020}  \at{First-principles demonstration of
  diffusive-advective particle acceleration in kinetic simulations of
  relativistic plasma turbulence}.  \jt{Astrophys. J.}
  \bvol{893}~(1),  \pg{L7}.

\bibitem[{Yuan} {\em et~al.\/}(2016){Yuan}, {Nalewajko}, {Zrake}, {East} \&
  {Blandford}]{Yuan_2016}
{\sc \au{{Yuan}, Y.}, \au{{Nalewajko}, K.}, \au{{Zrake}, J.}, \au{{East},
  W.~E.} \& \au{{Blandford}, R.~D.}} \yr{2016}  \at{{Kinetic Study of
  Radiation-reaction-limited Particle Acceleration During the Relaxation of
  Unstable Force-free Equilibria}}.  \jt{Astrophys. J.}  \bvol{828},  \pg{92}.

\bibitem[Zenitani \& Hoshino(2001)]{Zenitani_2001}
{\sc \au{Zenitani, S.} \& \au{Hoshino, M.}} \yr{2001}  \at{The generation of
  nonthermal particles in the relativistic magnetic reconnection of pair
  plasmas}.  \jt{Astrophys. J.}  \bvol{562}~(1),  \pg{L63--L66}.

\bibitem[Zenitani \& Hoshino(2007)]{Zenitani_2007}
{\sc \au{Zenitani, S.} \& \au{Hoshino, M.}} \yr{2007}  \at{Particle
  acceleration and magnetic dissipation in relativistic current sheet of pair
  plasmas}.  \jt{Astrophys. J.}  \bvol{670}~(1),  \pg{702--726}.

\bibitem[Zhdankin {\em et~al.\/}(2017{\natexlab{{\em a\/}}})Zhdankin, Uzdensky,
  Werner \& Begelman]{10.1093/mnras/stx2883}
{\sc \au{Zhdankin, V.}, \au{Uzdensky, D.~A.}, \au{Werner, G.~R.} \&
  \au{Begelman, M.~C.}} \yr{2017{\natexlab{{\em a\/}}}}  \at{{Numerical
  investigation of kinetic turbulence in relativistic pair plasmas – I.
  Turbulence statistics}}.  \jt{Mon. Not. R. Astron. Soc.}  \bvol{474}~(2),  \pg{2514--2535}.

\bibitem[Zhdankin {\em et~al.\/}(2018)Zhdankin, Uzdensky, Werner \&
  Begelman]{Zhdankin_2018}
{\sc \au{Zhdankin, V.}, \au{Uzdensky, D.~A.}, \au{Werner, G.~R.} \&
  \au{Begelman, M.~C.}} \yr{2018}  \at{System-size convergence of nonthermal
  particle acceleration in relativistic plasma turbulence}.  \jt{Astrophys. J.}  \bvol{867}~(1),  \pg{L18}.

\bibitem[Zhdankin {\em et~al.\/}(2019)Zhdankin, Uzdensky, Werner \&
  Begelman]{PhysRevLett.122.055101}
{\sc \au{Zhdankin, V.}, \au{Uzdensky, D.~A.}, \au{Werner, G.~R.} \&
  \au{Begelman, M.~C.}} \yr{2019}  \at{Electron and ion energization in
  relativistic plasma turbulence}.  \jt{Phys. Rev. Lett.}  \bvol{122},
  \pg{055101}.

\bibitem[Zhdankin {\em et~al.\/}(2020)Zhdankin, Uzdensky, Werner \&
  Begelman]{10.1093/mnras/staa284}
{\sc \au{Zhdankin, V.}, \au{Uzdensky, D.~A.}, \au{Werner, G.~R.} \&
  \au{Begelman, M.~C.}} \yr{2020}  \at{{Kinetic turbulence in shining pair
  plasma: intermittent beaming and thermalization by radiative cooling}}.
  \jt{Mon. Not. R. Astron. Soc.}  \bvol{493}~(1),
  \pg{603--626}.

\bibitem[Zhdankin {\em et~al.\/}(2017{\natexlab{{\em b\/}}})Zhdankin, Werner,
  Uzdensky \& Begelman]{PhysRevLett.118.055103}
{\sc \au{Zhdankin, V.}, \au{Werner, G.~R.}, \au{Uzdensky, D.~A.} \&
  \au{Begelman, M.~C.}} \yr{2017{\natexlab{{\em b\/}}}}  \at{Kinetic turbulence
  in relativistic plasma: From thermal bath to nonthermal continuum}.
  \jt{Phys. Rev. Lett.}  \bvol{118},  \pg{055103}.

\bibitem[Zrake(2016)]{Zrake_2016}
{\sc \au{Zrake, J.}} \yr{2016}  \at{{Crab} {flares} {due} {to} {turbulent}
  {dissipation} {of} {the} {pulsar} {striped} {wind}}.  \jt{Astrophys. J.}  \bvol{823}~(1),  \pg{39}.

\bibitem[Zrake \& Arons(2017)]{Zrake_2017}
{\sc \au{Zrake, J.} \& \au{Arons, J.}} \yr{2017}  \at{Turbulent magnetic
  relaxation in pulsar wind nebulae}.  \jt{Astrophys. J.}
  \bvol{847}~(1),  \pg{57}.

\bibitem[Zrake \& East(2016)]{Zrake_East_2016}
{\sc \au{Zrake, J.} \& \au{East, W.~E.}} \yr{2016}  \at{{Freely} {decaying}
  {turbulence} {in} {force}-{free} {electrodynamics}}.  \jt{Astrophys. J.}  \bvol{817}~(2),  \pg{89}.

\end{thebibliography}

\end{document}